\documentclass{aastex63}

\def\M500{M_{500c}}

\def\r500{r_{500c}}

\def\Msun{\ensuremath{M_{\odot}}}

\def\sun{\ensuremath{_{\odot}}}
\def\Mstar{\ensuremath{M_{\star}}}

\newcommand\raisepunct[1]{\,\mathpunct{\raisebox{0.5ex}{#1}}}

\accepted{ApJ}

\watermark{Draft version \today}

\begin{document}

\title{Stellar mass measurements in Abell 133 with Magellan / IMACS}

\author{S.~Starikova}
\affiliation{Center for Astrophysics | Harvard \& Smithsonian \\
60 Garden Street, Cambridge, MA 02138, USA}
\author{A.~Vikhlinin}
\affiliation{Center for Astrophysics | Harvard \& Smithsonian \\
60 Garden Street, Cambridge, MA 02138, USA}
\author{A.~Kravtsov}
\affiliation{Department of Astronomy \& Astrophysics, \\
Kavli Institute for Cosmological Physics, \\ 
Enrico Fermi Institute, The University of Chicago \\
Chicago, IL 60637, USA}
\author{R.~Kraft}
\affiliation{Center for Astrophysics | Harvard \& Smithsonian \\
60 Garden Street, Cambridge, MA 02138, USA}
\author{T.~Connor}
\affiliation{The Observatories of the Carnegie Institution for Science \\
813 Santa Barbara Street, Pasadena, CA 91101, USA}
\affiliation{Jet Propulsion Laboratory, California Institute of Technology \\
4800 Oak Grove Drive, Pasadena, CA 91109, USA}
\author{J.~S.~Mulchaey}
\affiliation{The Observatories of the Carnegie Institution for Science \\
813 Santa Barbara Street, Pasadena, CA 91101, USA}
\author{D.~Nagai}
\affiliation{Department of Physics, Yale University \\
New Haven, CT 06520, USA}
\affiliation{Department of Astronomy, Yale University \\ 
New Haven, CT 06520, USA}
\affiliation{Yale Center for Astronomy \& Astrophysics, Yale University \\
New Haven, CT 06520, USA}

\shortauthors{Starikova et al.}
\shorttitle{Stellar mass measurements in A133}

\correspondingauthor{Svetlana Starikova, Alexey Vikhlinin}
\email{svetlana.starikova@cfa.harvard.edu, avikhlinin@cfa.harvard.edu}

\begin{abstract}
  We present the analysis of deep optical imaging of the galaxy
  cluster Abell 133 with the IMACS instrument on Magellan. Our
  multi-band photometry enables stellar mass measurements in the
  cluster member galaxies down to a mass limit of
  $M_\star=3\times10^8\,M_\odot$ ($\approx 0.1$ of the Large
  Magellanic Cloud stellar mass). We observe 
a clear difference in the
  spatial distribution of large and dwarf galaxies within the
  cluster. Modeling these galaxies populations separately, we can
  confidently track the distribution of stellar mass locked in the
  galaxies to the cluster's virial radius. The extended envelope of
  the cluster's brightest galaxy can be tracked to $\sim
  200$\,kpc. The central galaxy contributes $\sim 1/3$ of the the
  total cluster stellar mass within the radius $\r500$.
\end{abstract}


\section{Introduction}
\label{sec:introduction}

Clusters of galaxies have total gravitating masses of $\sim
10^{14}-10^{15}\ M_\odot$ and are the most massive systems that have
had time to collapse in the standard $\Lambda$CDM cosmology
\citep[see][for a review]{2012ARA&A..50..353K}. Given a mean comoving
density of matter in the universe of $\approx 4\times 10^{10}\,
M_\odot\,\rm Mpc^{-3}$, the large cluster masses imply that their
matter was assembled from regions of $\sim 15-50\,\rm Mpc$ in
size. Although clusters tend to form in high-density regions
\citep{1984ApJ...284L...9K}, the vast scales involved in their
formation means that, at least within roughly the virial radius, the
enclosed matter should have a mix of baryons and dark matter close to
the universal value
\citep[e.g.,][]{1993Natur.366..429W,1999ApJ...525..554F,2003MNRAS.339.1117V,2004MNRAS.355.1091K,2005ApJ...625..588K}. Furthermore,
the sizes of their virialized regions after collapse are $R\sim 1-5$
Mpc and their binding energies, $GM^2/R$, are thus $\sim
10^{63}-10^{64}$ ergs. Therefore, even the most energetic Active
Galactic Nuclei (AGN) feedback cannot eject baryons from deep
potential wells of clusters and significantly lower their baryon mass
fraction
\citep[e.g.,][]{2013MNRAS.431.1487P,2013ApJ...777..123B,2018MNRAS.479.5385H}. This
means that clusters should be approximately closed systems.

Studies of the total baryon fractions in clusters can thus be used as
a valuable test of the overall structure formation paradigm
\citep[see, e.g.,][for a review]{2011ARA&A..49..409A}. At the same
time, baryon mass fractions within the radii readily accessible by
current X-ray observations, $\sim 0.5$ of the virial radius, are well
below the values of the universal baryon fraction derived from Cosmic
Microwave Background fluctuations
\citep{2006ApJ...640..691V,2012ApJ...745L...3L,2016A&A...592A..12E},
and this is yet to be fully explained by cosmological cluster
simulations \citep[e.g.,][]{2017MNRAS.471.1088B}.  Furthermore, the
distribution of stellar material and hot gas within clusters should
bear the imprint of key processes shaping galaxy formation. Indeed,
observed stellar mass fractions and stellar mass function of cluster
galaxies have become valuable benchmarks for testing models of
feedback in cosmological simulations of cluster formation
\citep[e.g.,][]{2014MNRAS.443.1500M,2016MNRAS.459.4408M,2017MNRAS.470.4186B,2017MNRAS.465.2936M,2018MNRAS.480.2898C,2018MNRAS.475..648P,2018MNRAS.479.5385H}. The
radial profile of stellar density of the Brightest Cluster Galaxy
(BCG), as well as the radial distribution of stellar mass in galaxies are
potentially equally powerful constraints on the models
\citep[e.g.,][]{2014MNRAS.443.1500M,2018MNRAS.476.4543B}.

Despite recent progress
\citep{2013ApJ...778...14G,2012MNRAS.423..104B,2014MNRAS.437.1362B,2018AstL...44....8K,2018MNRAS.475.3348H},
the number of clusters with available accurate measurements of the gas
mass, stellar mass in galaxies down to dwarf scales, and stellar
material in the outer envelope of the central galaxy remains small.
The main goal of this study is to accurately measure the contribution
of the stellar populations (individual galaxies and intra-cluster
light) to the total baryon budget in the cluster Abell 133.

Abell 133 is a massive nearby ($z=0.05695$) galaxy cluster with
extensive mapping of surrounding distribution of galaxies and
filamentary cosmic web structure
\citep{2018ApJ...867...25C,2019arXiv190910518C}, as well as deep X-ray
observations by the {\sl Chandra} X-ray Observatory \citep[][Vikhlinin
et al., in
preparation]{2006ApJ...640..691V,2013HEAD...1340101V,2014MNRAS.437.1909M}. The
cluster has a cool core and prominent radio relics indicative of the
ongoing merger activity \citep{2010ApJ...722..825R}, although
distribution of galaxies does not reveal clear signs of dynamical
disturbance \citep{2018ApJ...867...25C}. Hydrostatic equilibrium
analysis using X-ray {\sl Chandra} observations give total mass within
the radius enclosing density contrast equal to the 500 times the
critical density at the redshift of the cluster of $\M500\approx
3.42\times 10^{14}\,M_\odot$ and corresponding radius
$\r500=1.048\,\rm Mpc$.

The paper is organized as follows. In Section \ref{sec:observations}
we explain the choice of fields within A133 and neighboring fields
used to estimate background galaxy density, describe observations of
these fields using the IMACS camera on the 6.5m Magellan Baade
telescope and present their images and general discussion of the
features they reveal. 

In Section \ref{sec:data} we describe data reduction procedures,
assumptions, and methods we use to carry out source detection,
classification, galaxy photometry, and sample completeness. In Section
\ref{sec:mstar-function} we describe the method we use to estimate
stellar masses from galaxy luminosities and colors (\S
\ref{sec:galaxy-stellar-mass}) and results pertaining to the stellar
mass function (\S \ref{sec:smf}). In Section \ref{sec:mprof} we
present the radial distribution of galaxies of different stellar mass (\S
\ref{subsec:clust-memb-galax}), the radial stellar mass distribution
of the BCG (\S \ref{subsec:bcg}) and the total mass profile of all
stars in the cluster (\S \ref{subsec:total}). We discuss our findings
and their interpretation in Section \ref{sec:discussion} and summarize
our main results and conclusions in Section \ref{sec:conclusions}.

All distant-dependent quantities throughout this paper are computed assuming the nominal best-fit cosmological parameters from \cite{2014ApJ...794..135B}: $h=0.696$, $\Omega_{\rm M}=0.286$, and $\Omega_{\rm \Lambda}=0.714$. Galaxy luminosities are computed in the rest-frame using the Vega magnitude system.

\section{Observations}      
\label{sec:observations}

\def\wvdecomp{\textsc{wvdecomp}}

\begin{figure}
\centerline{\includegraphics[width=0.99\linewidth]{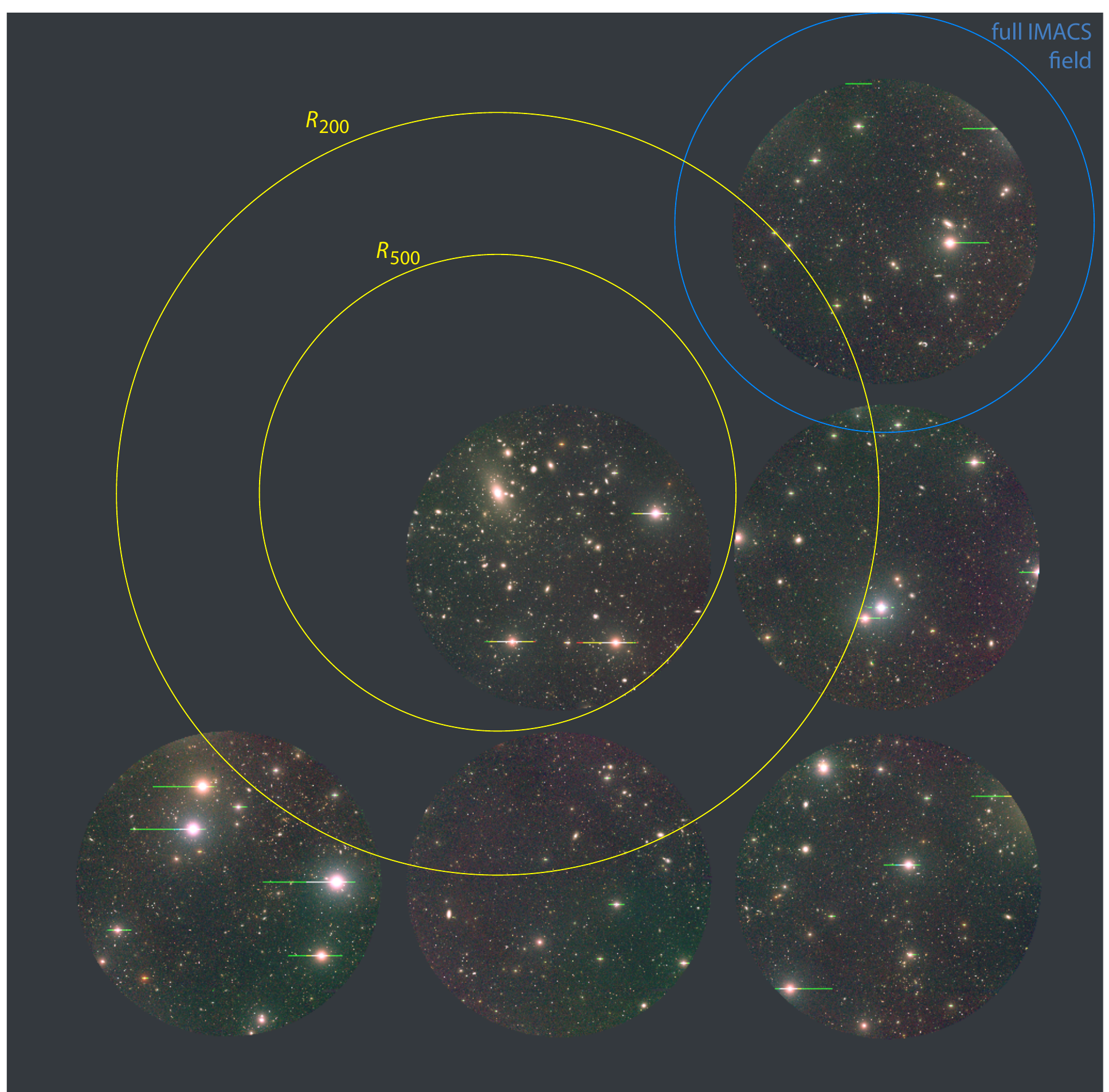}}
\caption{Mosaic of IMACS observations of A133, blending the images in
  the $R$, $V$, and $I$ filters. Only the inner 20~arcmin diameter
  regions of each field were used in the analysis (\S~\ref{sec:bg})
  and are shown here. The full IMACS field of view (27.4 arcmin
  diameter) is shown by the blue circle for guidance. Yellow circles
  mark $r_{500c}$ and $r_{200c}$ radii of the cluster. Note that all
  data shown in this image are the ``cluster'' fields. We have
  obtained an additional set of 8 background fields (not shown here)
  at $\sim 1.5^{\circ}$ off-cluster distances.}
\label{fig:cluster_fields}
\end{figure}

The galaxy cluster A133 was observed on the 6.5m Magellan Baade
Telescope over three nights in 2005. The observations were performed
with the IMACS \citep[Inamori-Magellan Areal Camera and
Spectrograph,][]{2011PASP..123..288D} instrument in its f/2 focus
configuration with the 8192$\times$8192 pixel Mosaic1 detector. The
central cluster region and South-East extension were covered with a
six-location grid (Figure~\ref{fig:cluster_fields}). The coverage
reaches outside the $r_{200c}$\footnote{$r_{200c}\simeq1.54r_{500c}$
  assuming concentration parameter indicated by the total mass profile
  of the cluster derived using hydrostatic equilibrium equation using
  X-ray data.} radius in the West, South, and South-West directions
from the cluster center. Unfortunately, the North-East corner of the
cluster remained unobserved. In analyzing the galaxy distributions
below, we make an assumption of azimuthal symmetry.

Since the fore- and background galaxy populations have to be
subtracted statistically in the cluster pointings, we obtained data
for their careful calibration. Specifically, we observed 8 background
fields at $\sim 1.5^{\circ}$ ($\approx 6\,$Mpc) distances from the
cluster center. Approximately half of all available exposure time was
spent in these background fields, so the background images reach the
same depth as the A133 pointings. Also, we constantly alternated
between the cluster and background fields during the night, so the
background pointings can be used for a measurement of the diffuse sky
background. This turned out to be crucial for analyzing the extended
diffuse light halo of the cluster central galaxy (see
\S~\ref{subsec:bcg} for details). We chose to observe the fields for
background estimation at $6\,\mathrm{Mpc}\approx3.5 r_{200c}$, as
opposed to using random pointings far away from the cluster, because
most of the volume at $z\sim 0$ is in low-density regions. At
the same time, galaxies and mass are strongly correlated with clusters
and the average profile of mass around clusters is expected to reach
mean density only at $\gtrsim 10R_{\rm vir}$
\citep[e.g.,][]{2014ApJ...789....1D}.  Significant contribution to the
relevant projected background is thus expected to be due to such
correlated structures relatively close to the cluster, while commonly
used estimates of the background using random fields will
underestimate the background \citep[see, e.g., discussion in \S 2.2.3
of][]{2017MNRAS.470.4767B}. Therefore, we chose to estimate
background at the radii well outside the virial radius, but still
sufficiently close to the cluster to give us a realistic estimate of
the background population.

All fields were observed in the Bessel $V$, Bessel $R$, and CTIO $I$ filters.
In each filter, several exposures were taken with $\sim 15''$ dither to
facilitate removal of cosmic rays and cosmetic defects of the
CCDs. The total exposure per location per filter was in the range from
300 to 1800 sec. The deepest images were taken in the $R$-band (typical
exposures $\sim 900$ sec), while $V$- and $I$-band images are
shallower (typically, 300 sec). The deepest exposures were taken for
the central cluster field and one of the background fields --- 1800,
1500 and 900 sec in $R$, $V$, and $I$ filters, respectively.  Seeing
varied during the observing run in the range $0.6''-1.2''$, but stayed
sub-arcsec for a large fraction of the $R$-band observations.  For
accurate photometric calibration, we observed the standard star field
SA~98 \cite{1992AJ....104..340L} in $R$, $V$, and $I$ filters in the
pre-dawn hours of each night.

When this paper was in preparation, the first release of the DES
\citep[Dark Energy Survey][]{2018ApJS..239...18A} data near the A133
location became available. The DES data cover a larger area around the
cluster and provide accurate photometry in the SDSS filters. However,
we find that the DES images are shallower than our Magellan data (see
Figure~\ref{fig:detections} below). We, therefore, used the DES
catalogs to verify the accuracy of photometric measurements and
stellar mass determinations from Magellan data for commonly detected
galaxies (see \S~\ref{sec:desVSmag}).

\subsection{General discussion of Magellan images}

The composite Magellan image (Figure~\ref{fig:cluster_fields}) clearly
shows a large number of A133 member galaxies. The cluster is dominated
by the brightest central galaxy. We show below that the BCG, including
its extended envelope, contributes over 30\% of the total stellar mass in the cluster
within $\r500$ and $\sim 50\%$ within $0.5\,\r500$ (\S~\ref{subsec:total}). There are a few other structures worthy of a brief discussion.

\begin{figure}
\centerline{\includegraphics[width=0.99\linewidth]{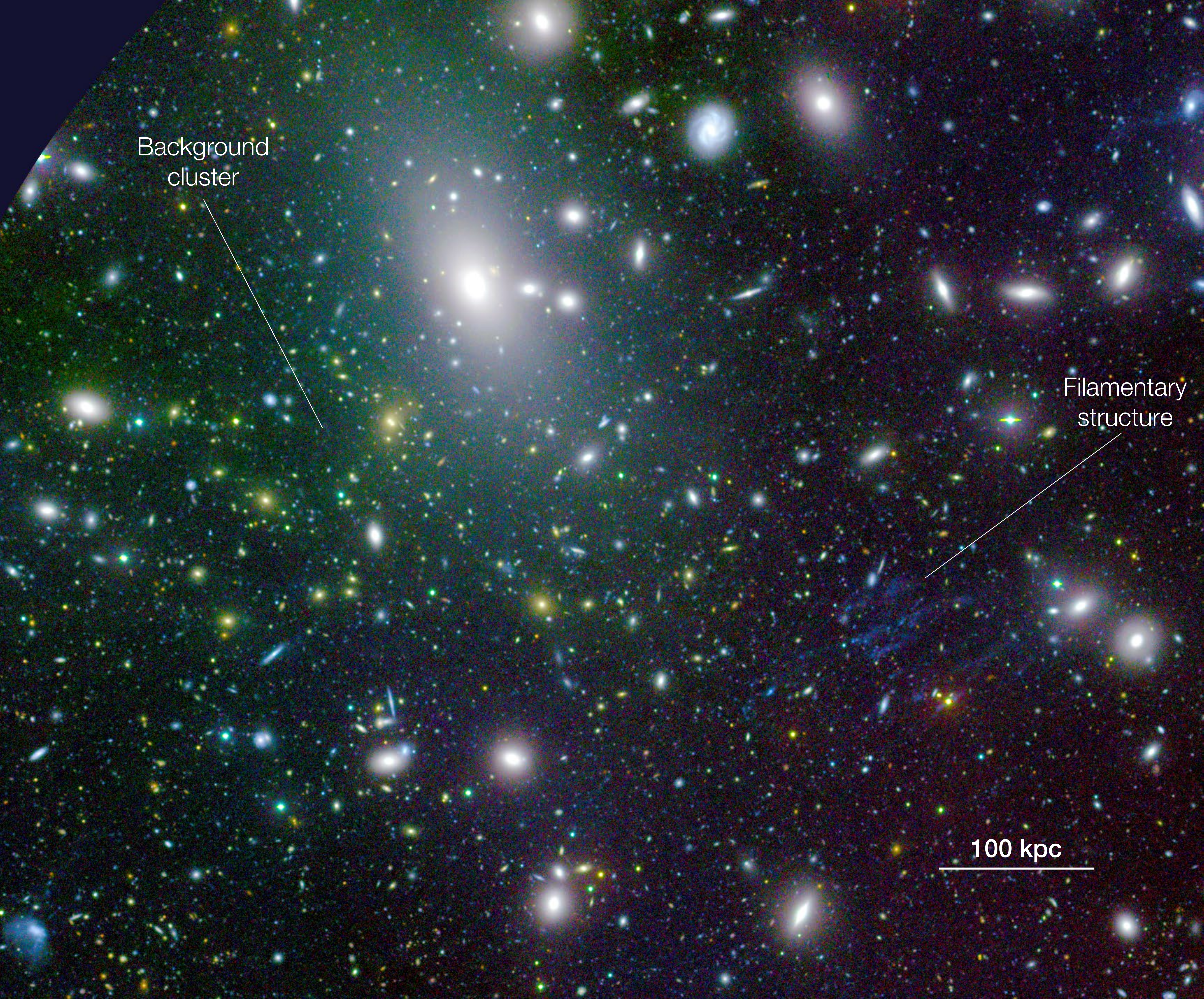}}
\caption{Zoom-in on the central region of A133. North is up, East is to the left, and the colors in the
  $V$, $R$, and $I$ blend are chosen so that the A133 red sequence
  corresponds to white. Approximately $\sim300$~kpc South-West of the
  BCG there is a filamentary structure with bluer colors than the
  population of A133 ellipticals. Approximately 150\,kpc to the
  South-East of the A133 BCG there lies a projected background cluster
  ($z=0.29$), apparent as a group of redder galaxies
  (\S~\ref{subsec:Red sequence}).}
\label{fig:filament}
\end{figure}

In Figure~\ref{fig:filament}, we show a zoom-in on the cluster central
region. The colors represent the relative flux in the $V$, $R$, and $I$
bands, and are chosen such the color of galaxies on the A133's red
sequence is white. In addition to a large number of A133 members, the
image clearly shows a concentration of fainter, red galaxies
$\sim150$~kpc to the South-East of the BCG. This group of galaxies
forms a separate red sequence and corresponds to a background galaxy
cluster at $z\approx0.29$ (see \S~\ref{subsec:Red sequence} below).

At a distance of $\sim300$~kpc to the South-West of the BCG, there is
a filamentary structure with bluer colors than the A133 elliptical
galaxies. This structure possibly corresponds to a tidally disrupted
cluster member. It has a size of $~100$~kpc, apparent magnitude $R =
19.8$~mag, and color $V-R=0.19$~mag, $\approx 0.3$~mag bluer than the
A133 red sequence (c.f.\ Figure~\ref{fig:color-mag} below). Other
examples of tidally disrupted galaxies in nearby clusters have been
reported, such as UGC~6697 in A1367 \citep{2005ApJ...621..718S} and
ESO~137-001 in A3627 \citep{2007ApJ...671..190S}. These objects show
$\sim50$~kpc tails of H$\alpha$ and X-ray emission. The filamentary
structure in A133 is larger in size, and does not show any X-ray tails
in the sensitive \emph{Chandra} data (no narrow-band H$\alpha$ imaging
is available at the time of this writing). Further analysis of these
structures will require additional data.

\section{Data Analysis}
\label{sec:data}

\subsection{Basic Image Reduction}

Basic data reduction steps including bias removal, dark current
correction, and flat fielding using twilight flats were performed with
IRAF's CCDPROC package. Individual images in each dither pattern were
merged using a combination of mean and median averaging with
sigma-clipping. This procedure automatically removes the cosmic rays
and cosmetic CCD defects.

The astrometric solutions were obtained using the
\emph{Astrometry.net} package \citep{2010AJ....139.1782L}. All images
were then resampled to a common tangential projections. This step was
necessary for accurate matching of the images obtained in individual
filters and for creating large-scale mosaics such as those shown in
Figure~\ref{fig:cluster_fields}. Because the IMACS f/2 camera was
slightly misaligned prior to adjustments made in 2006 and
2008\footnote{{\scriptsize\url{http://www.lco.cl/telescopes-information/magellan/instruments/imacs/user-manual/the-imacs-user-manual}}},
resampling to a global tangential projection resulted in small
aliasing effects. Those effects have not seriously affected the image quality
because the image pixel size, $0.2''$, is substantially smaller than
the seeing during the A133 observing run. However, aliasing modifies
the pixel-to-pixel noise in the final images, and has to be taken into
account during object detection (see \S~\ref{sec:detection} and
Appendix~\ref{sec:wvdecomp-and-noise} below).

\subsection{Background Subtraction}
\label{sec:bg}

To accurately calculate fluxes of bright and faint sources, we use
global and local background subtraction methods.  Sky brightness and
noise levels heavily varied among images of the same exposure in a
given field. However, we noticed that for every image, there was a
linear dependence between sky brightnesses in the center and the
center-to-edge difference. This dependence was exploited to remove the time
variations of the background. Therefore, we could build a
combined background image from all observed background fields taking
into account individual levels of noise and sky, and excluding point
sources. This image, the global background pattern, was subtracted from
all observations. Unfortunately, using this method we could not model
random spatial variations of the background which appear in some
fields. We applied local background subtraction method to produce
images which we used for detection and flux measurements for majority
of sources (see Appendix~\ref{sec:wvdecomp-and-noise} for a
description of this procedure).

Individually processed images were combined into the final master
images for every cluster and background field. The IMACS CCD camera
ideally delivers an 27.4 arcmin diameter field. However, there was
substantial loss of image quality (coma and astigmatism) near the edge
of the field of view prior to 2008, in addition to substantial
vignetting. We also found that the background pattern near the field
edge was unrepeatable, which was problematic for global background
subtraction. Therefore, we reduced the diameters of our fields to 20
arcmin.

  The bright, saturated stars render a portion of the cluster
  field unusable for detection of faint sources and accurate galaxy
  photometry. We masked out such regions ($\sim 40''$ radius) and
  excluded them from all further analysis. This excludes $\sim 3\%$ of
  the overall image area. More extended wings around bright stars are
  properly subtracted by our local background subtraction procedure.

\subsection{Source Detection \& Photometry}

Our general strategy was to detect sources and measure galaxy fluxes
in the $R$ band, and then measure fluxes at the same locations and
within the same apertures in the $V$ and $I$ bands. Our procedure, detailed
below, was designed to compensate for the difference in
sensitivity and seeing in different filters and cluster locations, and
to ensure accurate photometry even for very faint galaxies.

\begin{figure}
\centerline{\includegraphics[width=0.99\linewidth]{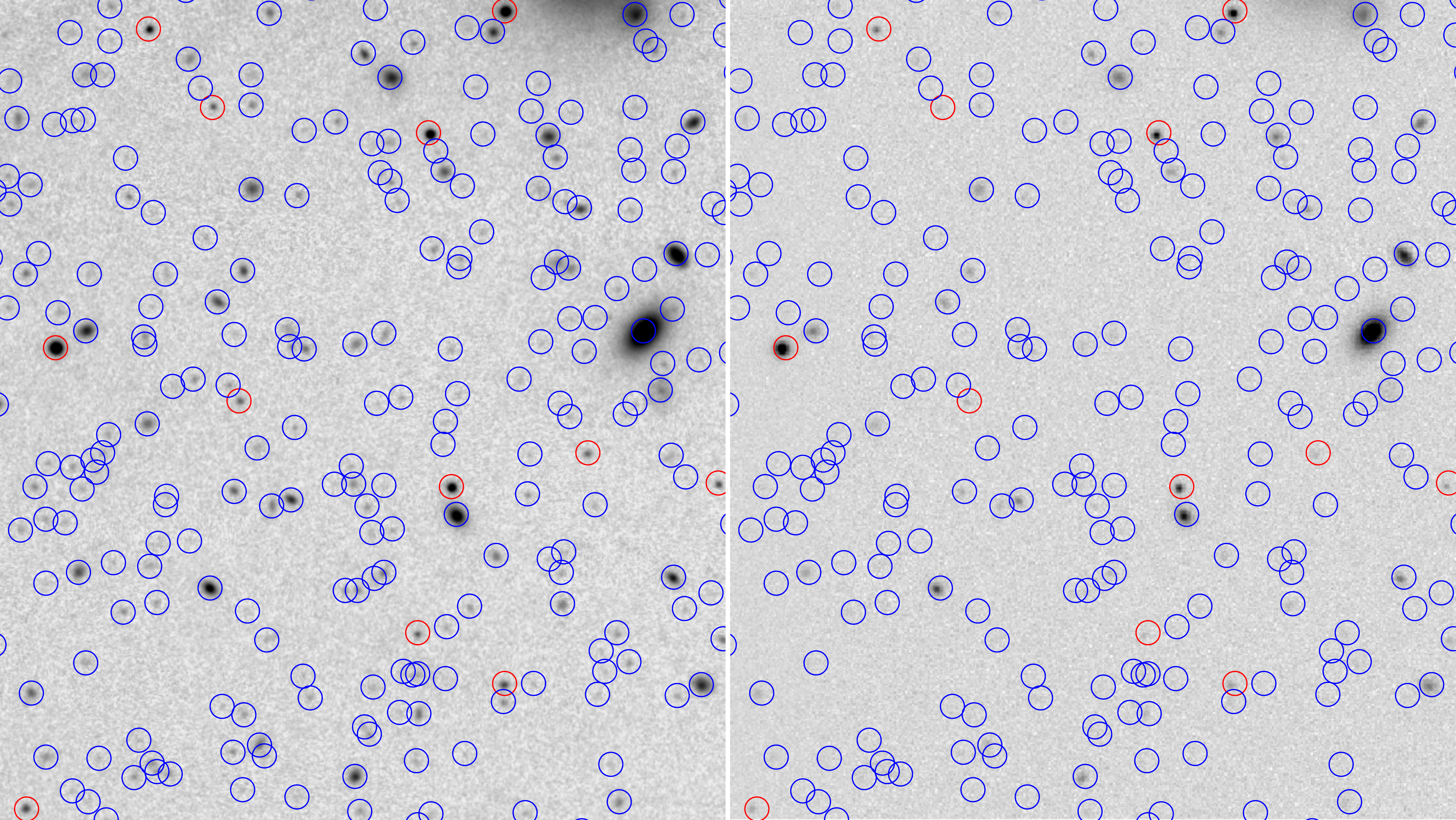}}
\caption{Illustration of the \wvdecomp-based detection in one of the
  cluster fields (circles). Blue circles are galaxies; red circles are
  stars identified by \emph{SExtractor}. The left panel shows our $R$-band
  Magellan image. The right panel shows the $r$-band DES image of the
  same region, with Magellan detections superimposed.}
\label{fig:detections}
\end{figure}

\subsubsection{Source Detection}
\label{sec:detection}

We start with running a source detection algorithm on the combined
$R$-band images. We used the wavelet decomposition algorithm,
\wvdecomp, which is proven to be very efficient for detection of faint
extended sources in X-ray images \citep{1998ApJ...502..558V}. We
computed the noise map to properly set detection thresholds at each
location. This is crucial for analyzing the faint galaxies in
  our Magellan images because pixel-to-pixel noise varies strongly
  within the field due to aliasing (see above) and non-uniform
  exposure coverage.
The noise
maps were empirically created from the data by convolving images
cleaned from sources with the \wvdecomp's wavelet kernel and
averaging the resulting rms deviations on $6''$ spatial scales (see
Appendix~\ref{sec:wvdecomp-and-noise} for details). \wvdecomp\ uses
this map for detection on the smallest scales; when proceeding to the
largest scales, the noise map is appropriately smoothed further by the
software \citep[see][for details]{1998ApJ...502..558V}.

The output from \wvdecomp\ is locations of statistically significant
sources (Figure~\ref{fig:detections}). We need a separate software
package to apply additional selection criteria and measure galaxy
fluxes. The first step is to identify and remove the likely stellar
sources. To this end, we have run the \emph{SExtractor}
\citep{1996A&AS..117..393B} detection on our $R$-band images,
cross-matched the \emph{SExtractor} and \wvdecomp\ source lists, and
removed sources for which \emph{SExtractor} measured stellarity
indices $>0.9$. On average $\approx 4\%$ of sources detected by
\wvdecomp\ were removed by this procedure.

\subsubsection{Fluxes and colors}
\label{subsec:fluxes and magnitudes}

Our main goal with the galaxy photometry is to reliably determine
\emph{total} luminosities for galaxies of different types and down to
low fluxes. We also need to ensure that the flux measurements are
consistent between exposures obtained in different filters and under
different seeing conditions. Our approach is as follows. We assume
that there are no color gradients within individual galaxies, as seems
to be the case for outer regions of massive spheroidal galaxies
\citep[see, e.g.,][]{2010AJ....140.1528L,2014MNRAS.443.1433D}. We fit
the observed $R$-band (best-exposed filter) surface brightness
profiles of each galaxy with an analytic model that includes the PSF
effects. This analytic fit is used to define the circular
aperture size for subsequent flux measurement and determine the
aperture correction. The apertures are chosen such that they are
reasonably small to ensure good signal-to-noise in the flux
measurements. At the same time, they are sufficiently large such that
the differences in seeing between different nights and filters lead to
negligible changes in the aperture correction factors.

The galaxy profiles were extracted in circular annuli, centered on the
surface brightness peak determined by \wvdecomp. The annuli were of
a constant log-width ($\lg(r_{\mathrm{max}}/r_{\mathrm{min}})=0.15$),
with a maximim radius equal to 1.5 times the maximum distance from the
source within this source's ``island'' (see
Appendix~\ref{sec:wvdecomp-and-noise}). The profiles exclude other
sources detected in the vicinity by masking out the image pixels
falling within these other sources' islands and scaling the measured
flux in the partially-masked annulus appropriately.

Our analytic model is motivated by the results of
\cite{2013ApJ...764L..31K}, who showed that the stellar surface
density profiles of galaxies of different morphological types have
approximately similar shape at radii $r\gtrsim r_n$ and largely differ
at $r\lesssim r_n$, where $r_n$ is approximately a half mass radius of
stellar distribution. Within that radius, the profile of early-type
galaxies is approximately described by the de Vaucouleurs model, while
the late-type galaxies follow the exponential profile. We further note
that Sersic-type function, $I(r) \propto \exp(-7.669\,
(r/r_{0})^{\gamma})$, can describe both the de Vacouleurs model and
the exponential model, depending on the values of $r_0$ and
$\gamma$. The \cite{2013ApJ...764L..31K} results indicate that the
Sersic index, $\gamma$, is not constant with radius, but is changing
near the radius $r_{n}$. We can approximate this by replacing
$r/r_{0}$ with a function $z=y^{\alpha}(1+y^2)^{1-{\alpha}/2}$, where
$y=r/r_{0}$. In this case, the effective Sersic is $\gamma$ for $r\gg
r_{0}$, and $\gamma\,\alpha$ for $r\ll r_{0}$.

To account for the PSF effects, one ideally needs to convolve a 2D
light distribution with a Gaussian, and then convert the result back
to the 1D radial profile. This approach is very
computationally-intensive. We found that instead of the 2D
convolution, the PSF effects can be sufficiently accurately
approximated by multiplying the profile by $\mathrm{erf}(r/p)$ where
$p$ is a free parameter fit individually for each galaxy. At $r\gg p$,
this does not modify the profile, while at $r\ll p$, it introduces a
flattening. Qualitatively, these are precisely the modifications
expected from finite seeing.

\begin{figure}

{\large

\centerline{%
  \includegraphics[width=0.49\linewidth]{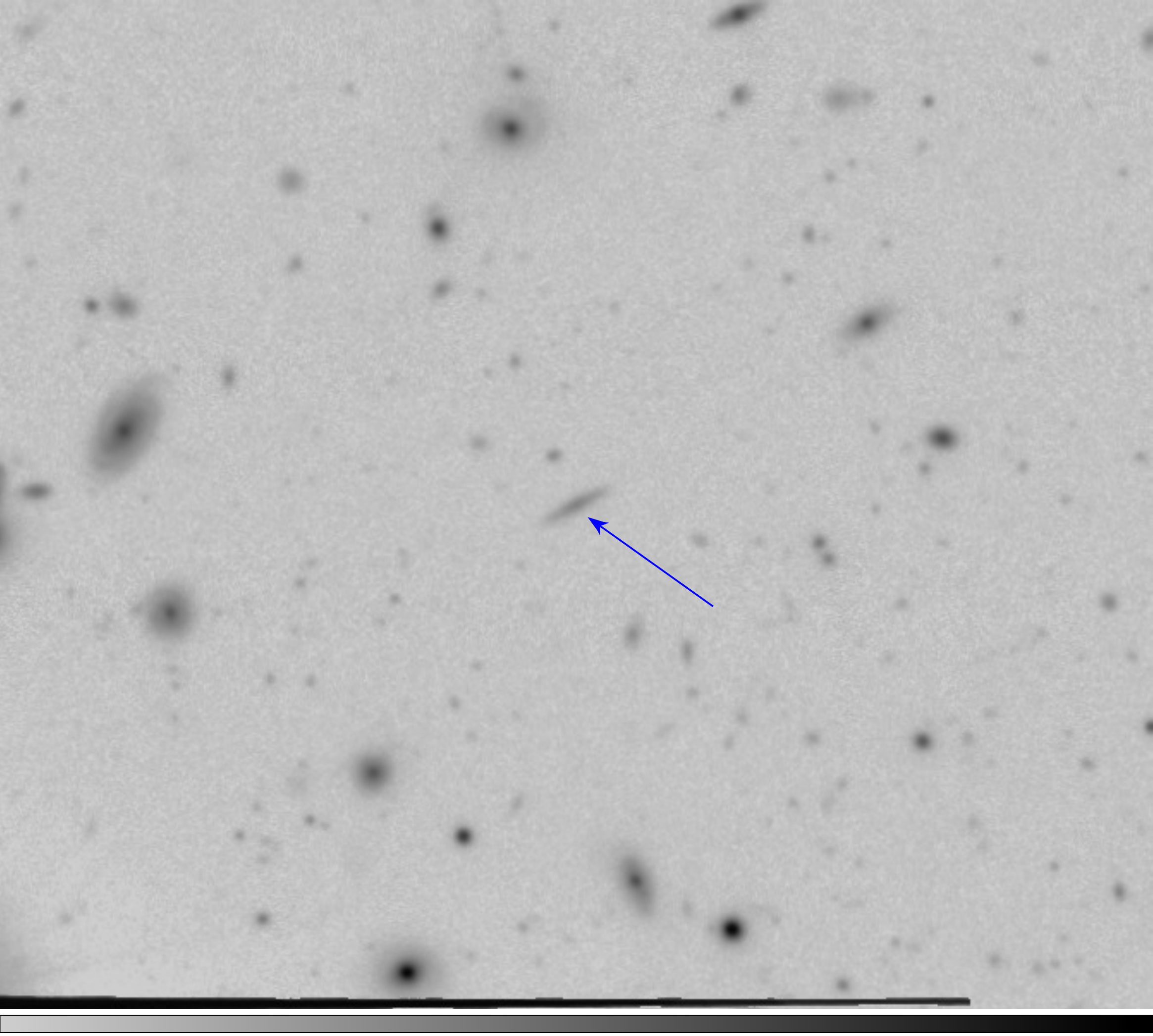}\hfill
  \includegraphics[width=0.49\linewidth]{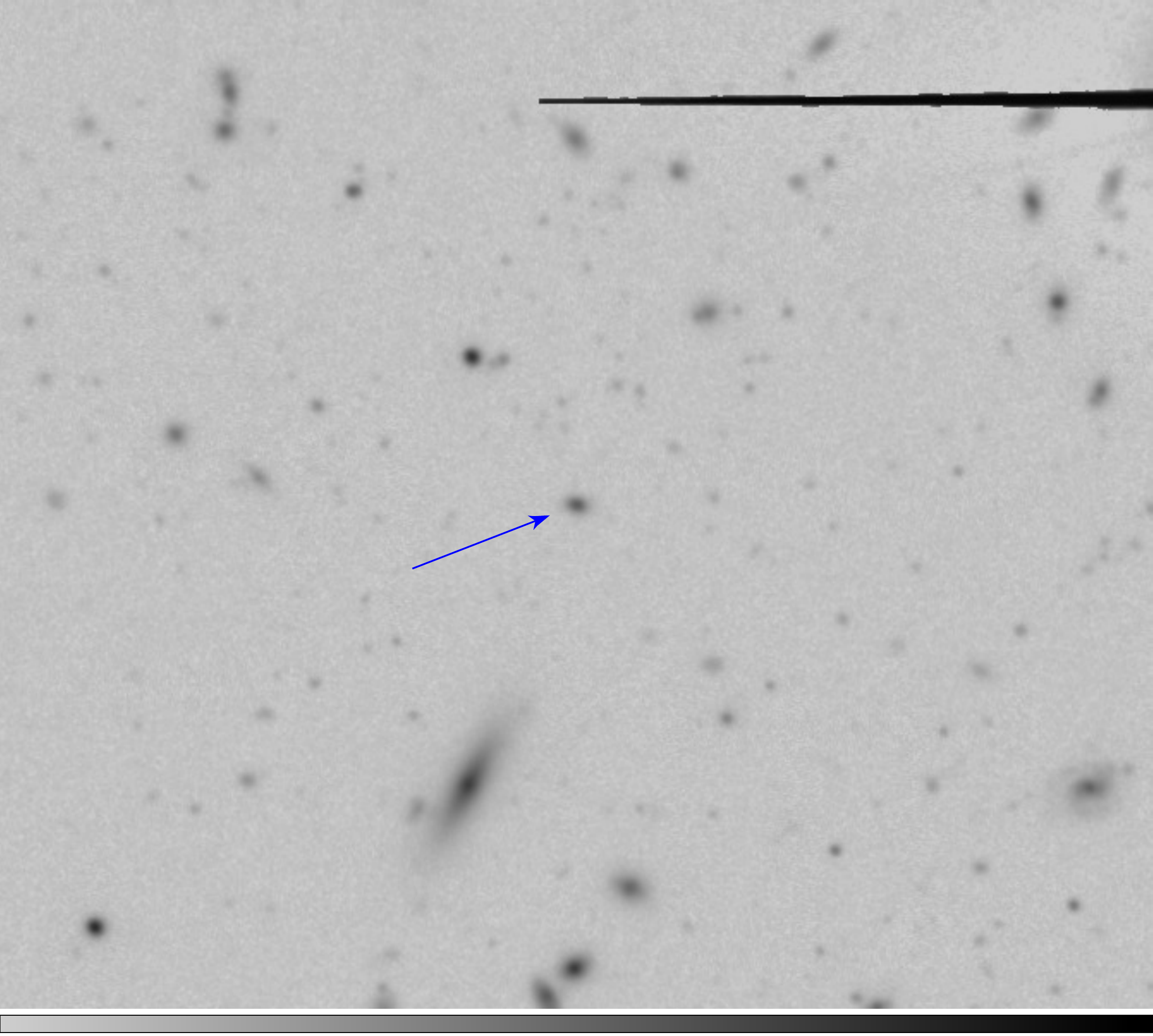}%
}
\vspace*{-0.428567\linewidth}
\vspace*{0.01\linewidth}

\noindent\hspace*{0.025\linewidth}%
\rlap{\emph{(a)}}%
\hspace*{0.51\linewidth}\emph{(b)}

\vspace*{-2\baselineskip}
\vspace*{-0.01\linewidth}
\vspace*{0.428567\linewidth}
\vspace*{0.04\linewidth}

\centerline{%
  \includegraphics[width=0.49\linewidth]{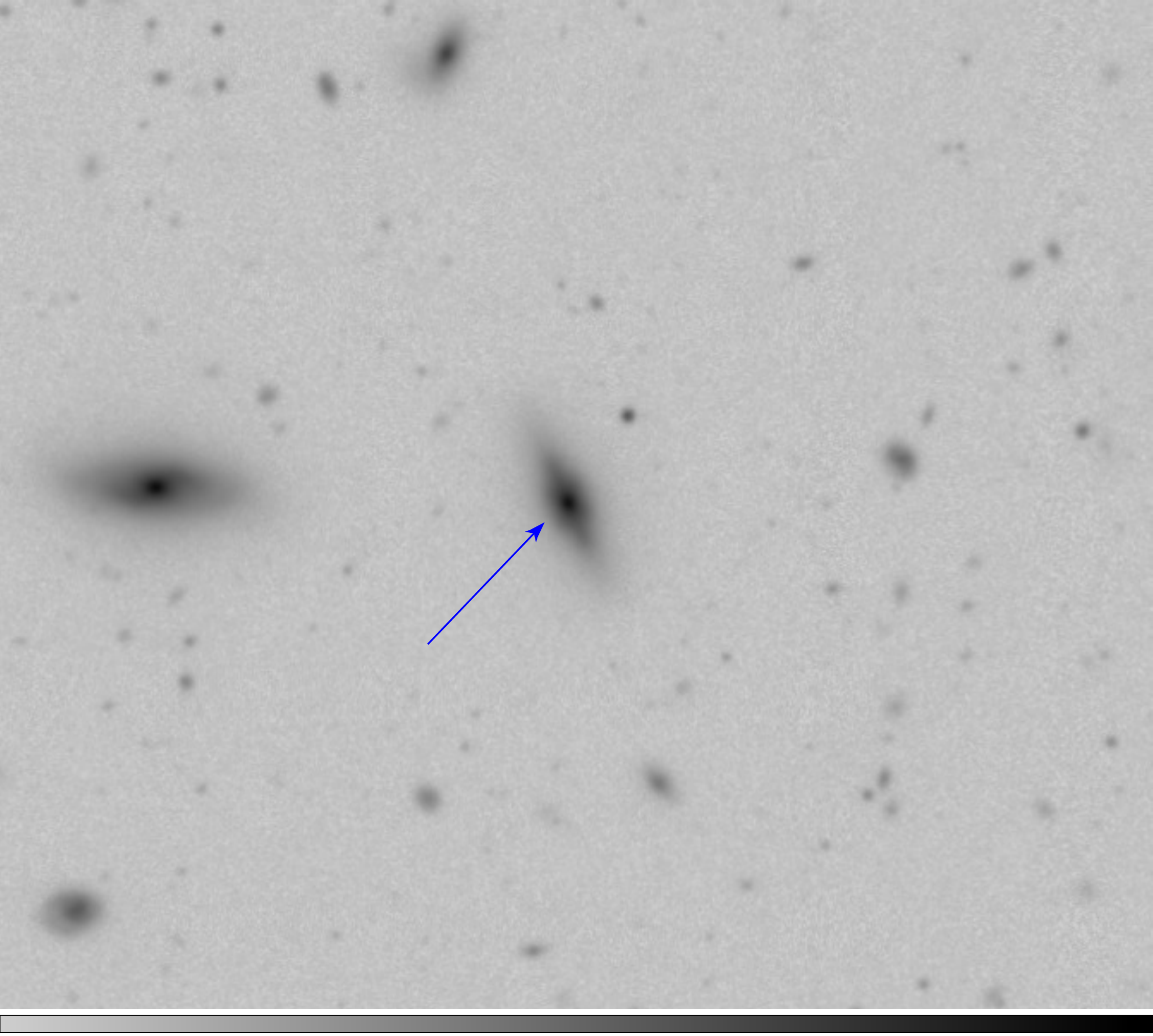}\hfill
  \includegraphics[width=0.49\linewidth]{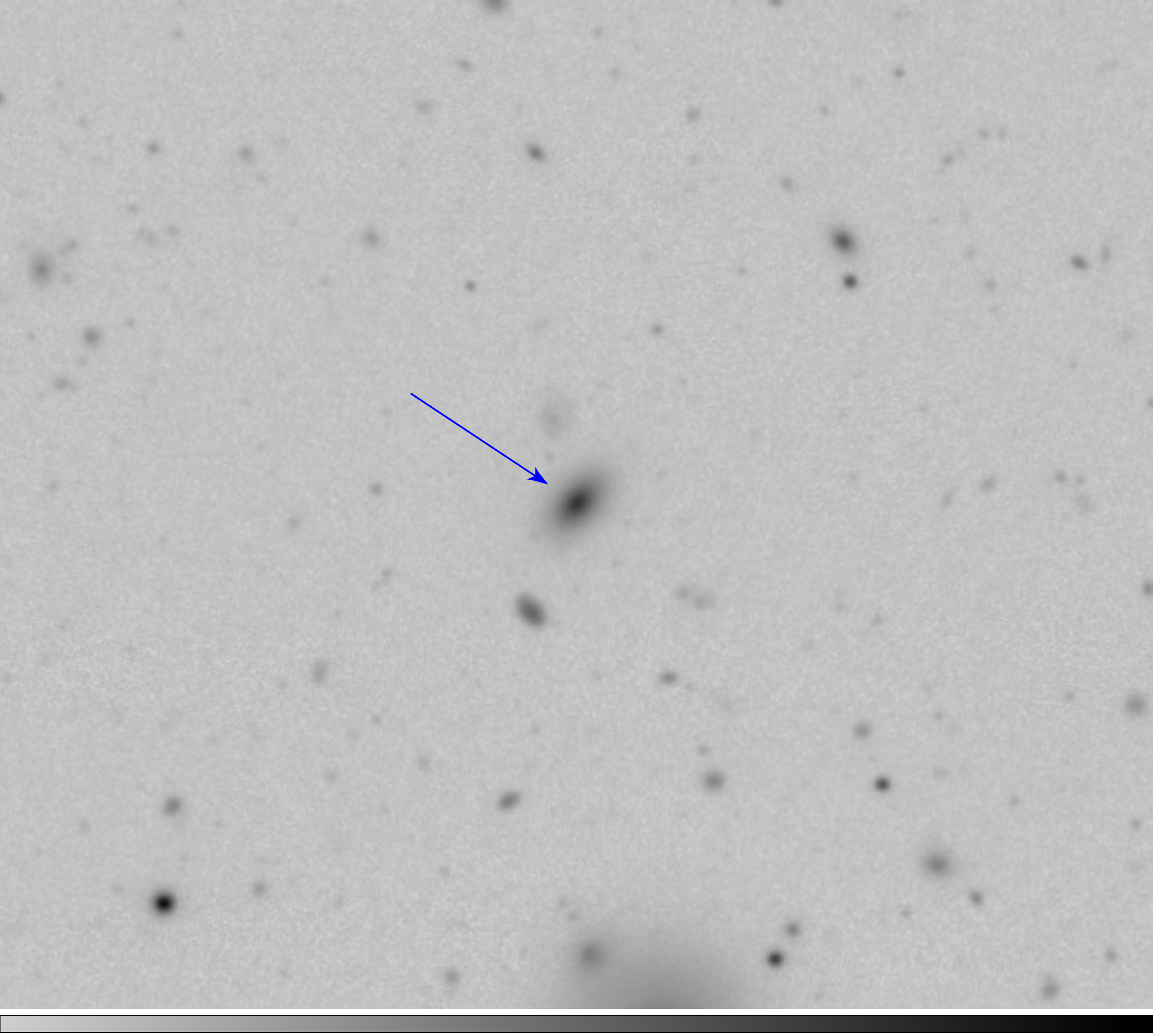}%
}

\vspace*{-0.428567\linewidth}
\vspace*{0.01\linewidth}

\noindent\hspace*{0.025\linewidth}%
\rlap{\emph{(c)}}%
\hspace*{0.51\linewidth}\emph{(d)}

\vspace*{-2\baselineskip}
\vspace*{-0.01\linewidth}
\vspace*{0.428567\linewidth}
\vspace*{0.02\linewidth}

}


\caption{Examples of galaxies whose profiles and best-fit models are
  shown in Figure~\ref{fig:profile:examples}. Examples include bright
  and faint spirals and ellipticals.}
\label{fig:profile:examples:images}
\end{figure}

\begin{figure}

{\large

\centerline{%
  \includegraphics[width=0.49\linewidth]{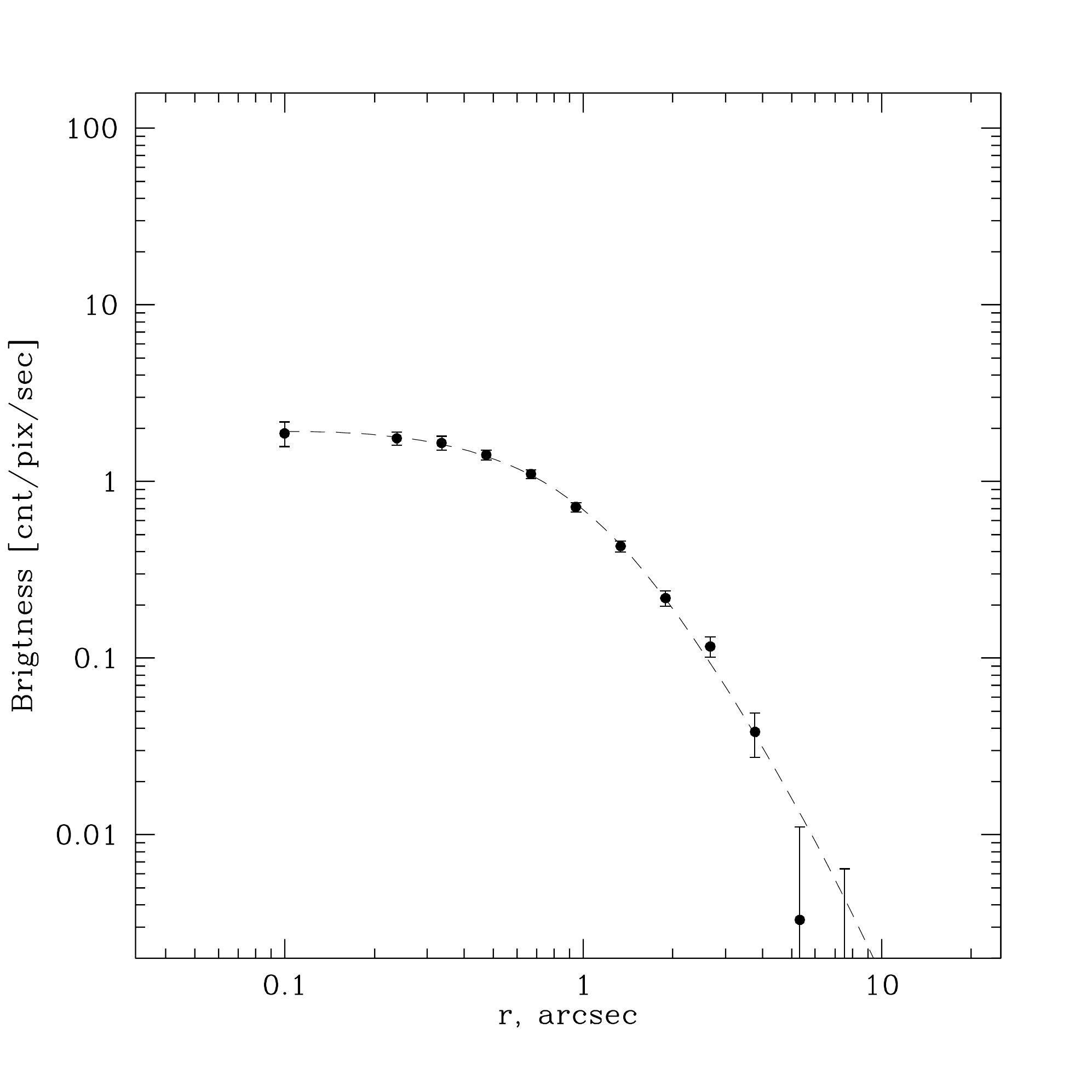}\hfill
  \includegraphics[width=0.49\linewidth]{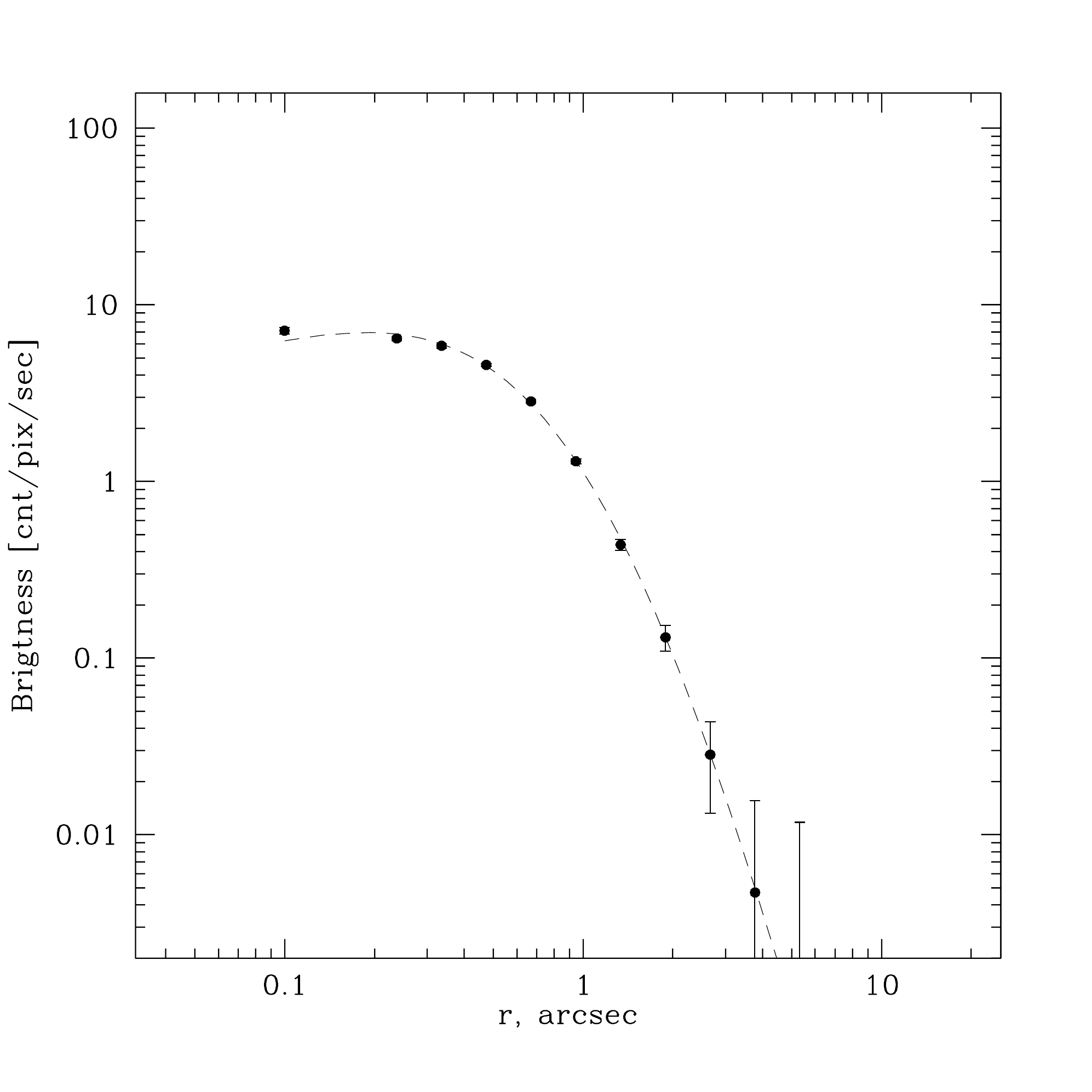}%
}
\vspace*{-0.49\linewidth}
\vspace*{0.06\linewidth}

\noindent\hspace*{0.39\linewidth}%
\rlap{\emph{(a)}}%
\hspace*{0.51\linewidth}\emph{(b)}

\vspace*{-2\baselineskip}
\vspace*{-0.06\linewidth}
\vspace*{0.49\linewidth}
\vspace*{0.04\linewidth}

\centerline{%
  \includegraphics[width=0.49\linewidth]{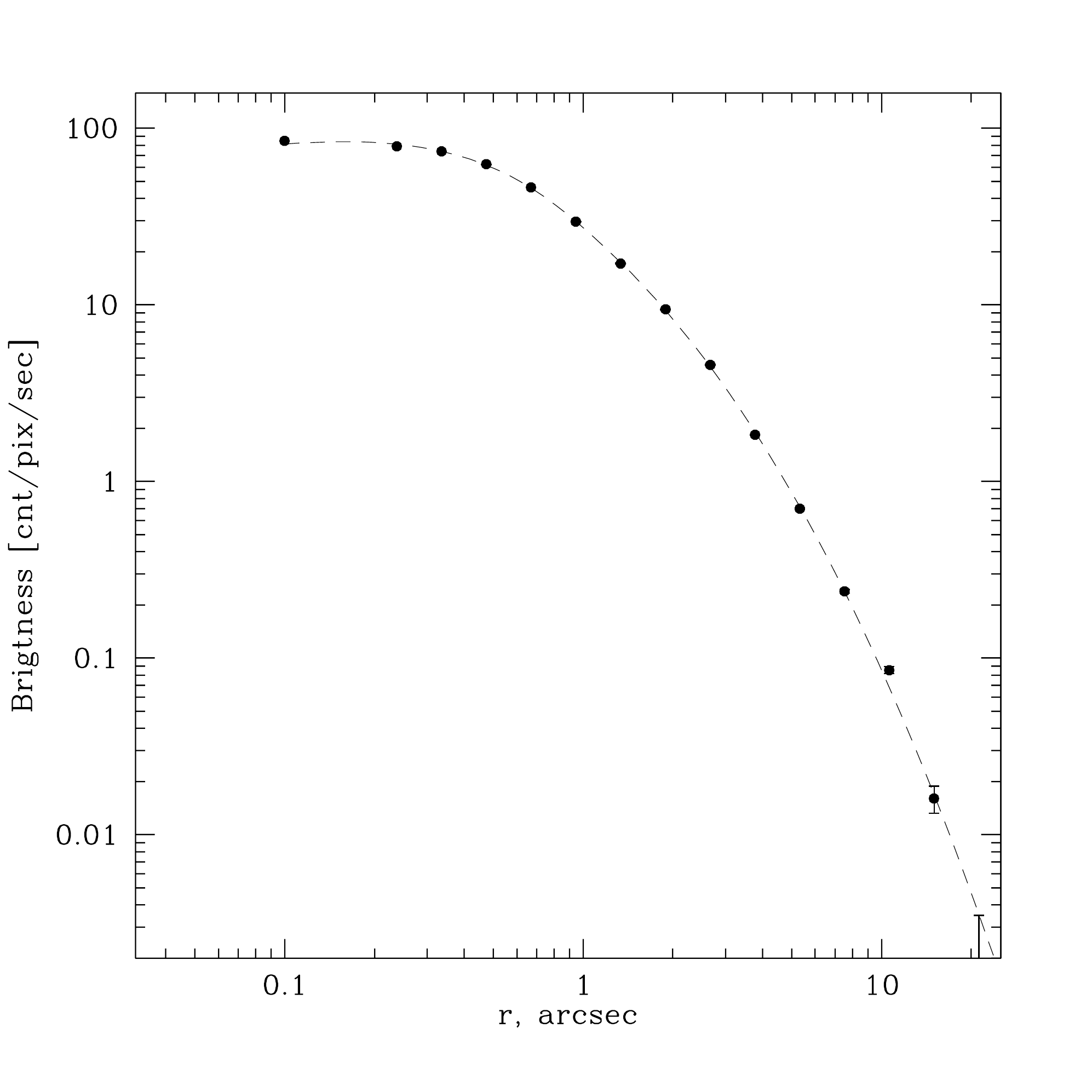}\hfill
  \includegraphics[width=0.49\linewidth]{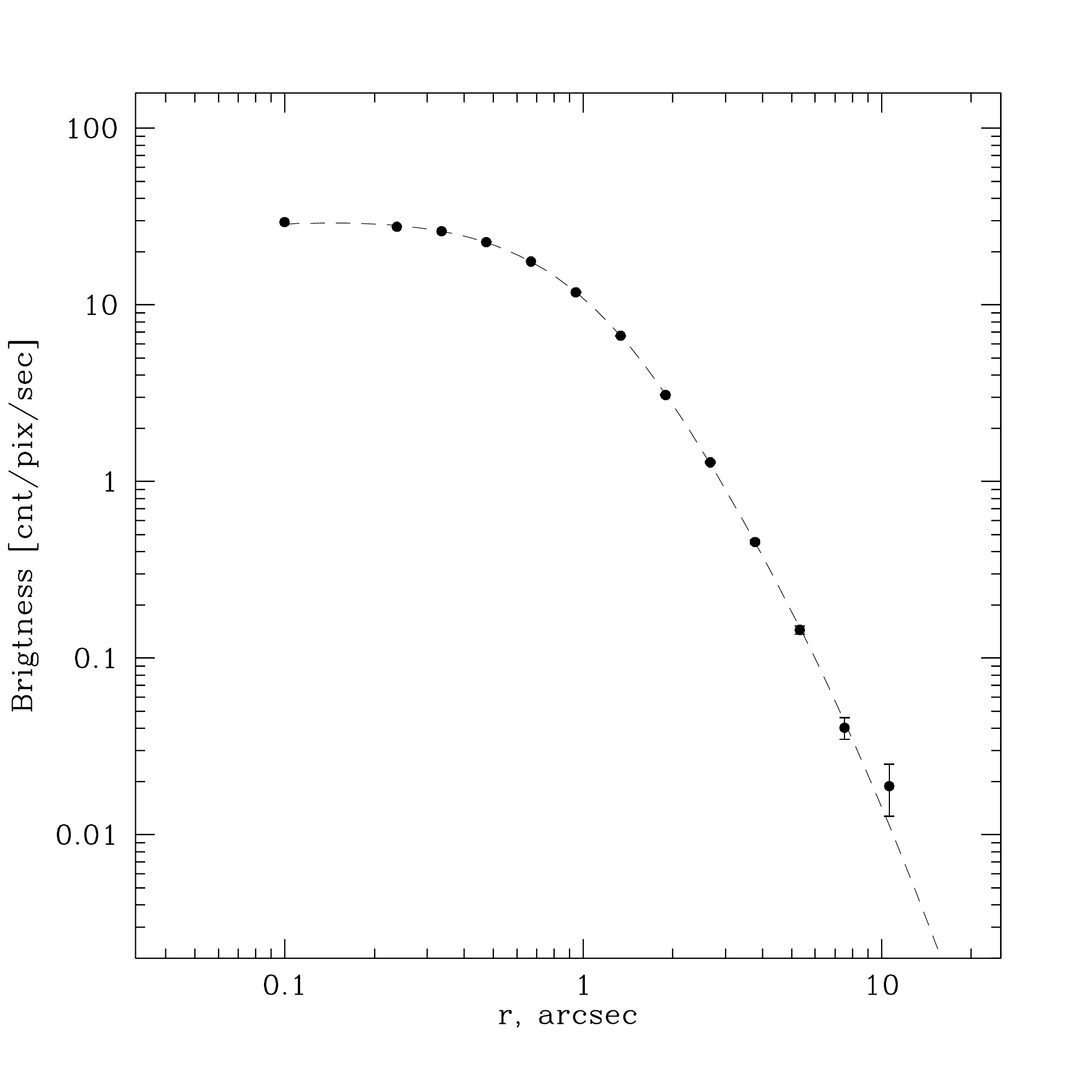}%
}

\vspace*{-0.49\linewidth}
\vspace*{0.06\linewidth}

\noindent\hspace*{0.39\linewidth}%
\rlap{\emph{(c)}}%
\hspace*{0.51\linewidth}\emph{(d)}

\vspace*{-2\baselineskip}
\vspace*{-0.06\linewidth}
\vspace*{0.49\linewidth}
\vspace*{0.02\linewidth}

}


\caption{Surface brightness profiles and best fit models
  (Eq.~\ref{frm:flux}) for the galaxies shown in
  Figure~\ref{fig:profile:examples:images}. These examples include
  both spirals (\emph{a} and \emph{c}) and ellipticals (\emph{b} and \emph{d}).}
\label{fig:profile:examples}
\end{figure}

To summarize, our analytic model is 
\begin{equation}
I(r) = I_0\exp(-7.669\,z^{\gamma})\,\times\,\mathrm{erf}(r/p) ,
\label{frm:flux}
\end{equation}
where $z=y^{\alpha}(1+y^2)^{1-{\alpha}/2}$, $y=r/{r_0}$, and $I_0,
r_0, \alpha, \gamma, p$ are model parameters. We found that this model
provides an accurate approximation to the data. The structural
parameters ($r_{0}$, etc.) can not be used literally because we treat
the PSF effects approximately. However, the total galaxy luminosities
can be obtained accurately, which is our goal. Examples of how this
model fits the profiles of typical spiral and elliptical galaxies in
our observations are shown in
Figure~\ref{fig:profile:examples:images}--\ref{fig:profile:examples}.

\begin{figure}
\centerline{\includegraphics[width=0.8\linewidth]{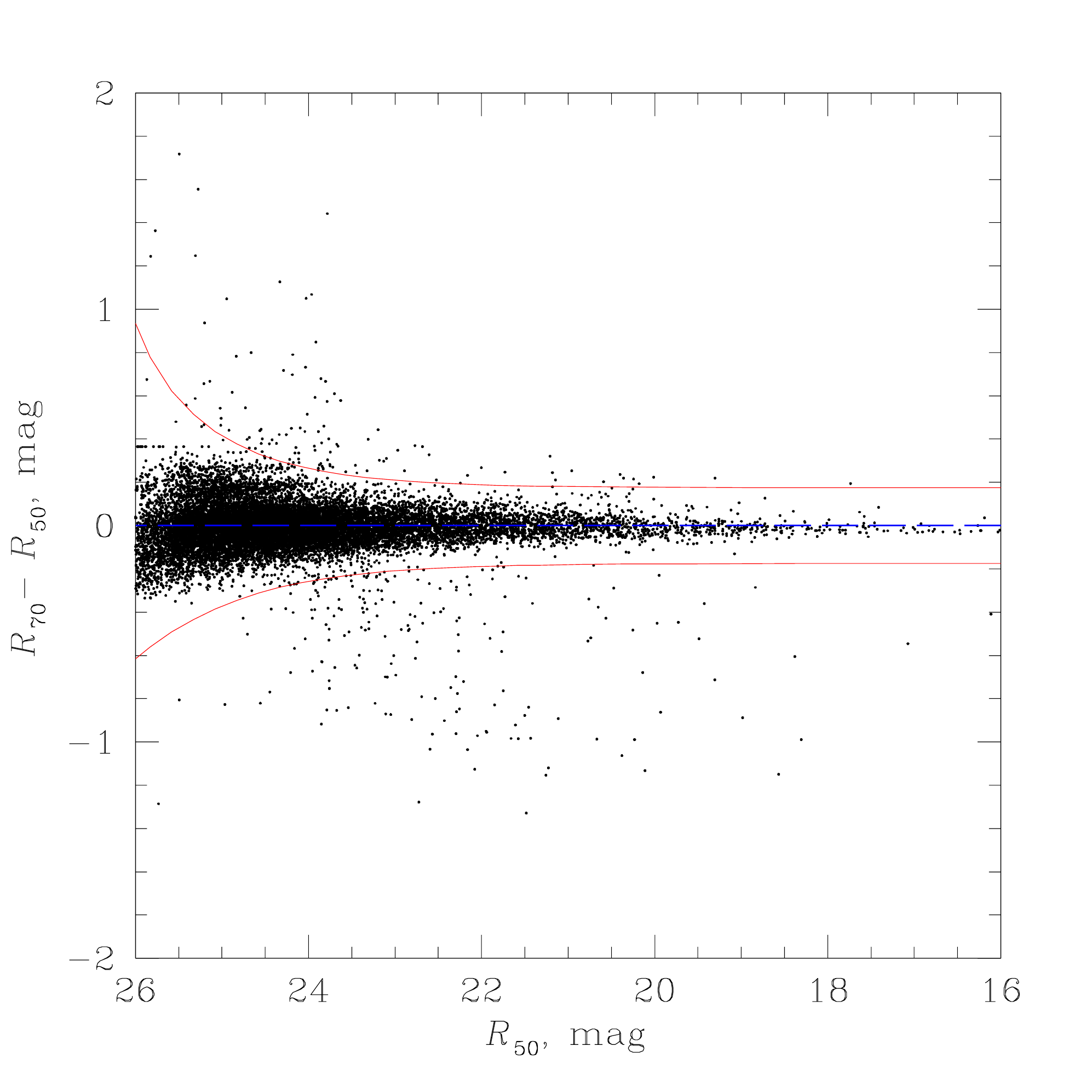}}
\caption{Comparison of the apparent $R$-band magnitudes of galaxies in
  the central cluster field, based on the aperture fluxes and
  correction factors for $r_{50}$ and $r_{70}$ radii (see text). For
  the majority of objects, there is an excellent agreement of the two
  measurements, with a small scatter shown by the solid red
  lines. There is a small percentage of outliers outside of the main
  relations indicated by red lines to guide the eye. We have
  hand-checked a large number of those. The outliers with the positive
  deviations of the $r_{70}$-based magnitudes were all found to be
  spurious detections in the wings of bright stars and galaxies. They
  were eliminated from further analysis. The systematic offset
    between the two fluxes is only $0.03$~mag.}
\label{fig:mag_f50f70}
\end{figure}

To check that the method for flux measurement is accurate and stable,
we performed the following test. The profile of Eq.\ \ref{frm:flux}
with parameters fit to the $R$-band images for each galaxy was used to
compute radii enclosing 50\%, 70\%, etc. of the total light. We then
measured the {\it actual\/} flux within these radii and estimated
total flux, as, e.g., $f_{\mathrm{tot}} = f_{50} / 0.5$ and compared
such estimate with the $f_{\mathrm{tot}}$ computed using the analytic
profile. If our model provides an accurate description of the observed
profiles, and the PSF effects are treated sufficiently accurately, the
two estimates of $f_{\mathrm{tot}}$ should agree and such comparison
thus represents a test of the accuracy of the model of eq
\ref{frm:flux}. In Figure~\ref{fig:mag_f50f70} we show an excellent
agreement of $f_{\mathrm{tot}}$ fluxes based on measurements in the
$r_{50}$ and $r_{70}$ radii. Since all aperture definitions work
equally well, we use the $r_{50}$-based fluxes in the further
analysis, to maximize signal-to-noise.

To compute colors for each galaxy, we calculate the fluxes in the
other filters using the apertures from the R-band, our deepest
images. The same aperture corrections are applied to all filters.
This is equivalent to an assumptions of no color gradients within
individual galaxies; the central cluster galaxy is the only object
where we accounted for the color gradients explicitly (see
\S~\ref{subsec:bcg} below). However, we need to apply an extra care in
selecting the aperture size, because the seeing for data obtained in
different filters can vary. We assume the PSF-related effects on the
galaxy brightness profiles are small outside the radius equal to the
FWHM of the PSF in that observation. Therefore, the galaxy aperture
was selected as the maximum of $r_{50}$ and the PSF FWHM's for
observations of the given field in $R$, $V$, and $I$ filters. After
the aperture size was determined this way, we computed the aperture
flux corrections using the best-fit model in the $R$-band. In
Figure~\ref{fig:logN-logS} below, we show a comparison of galaxy
number counts in different background fields, which shows an excellent
agreement above their completeness limits despite the seeing varying
from $0.6''$ to $1.2''$. This demonstrates that our modeling provides
a sufficiently accurate treatment of the PSF effects. A similar
excellent agreement in the source counts was found for the $V$- and
$I$-band data.

The automated flux measurement procedure described above is very
stable and works well for the vast majority of galaxies detected in
the IMACS images. The only objects, for which modifications were
needed, were bright, extended elliptical and spiral galaxies. For
bright ellipticals the main problem was that the locally measured
background (\S\,\ref{sec:bg}) over-subtracted the outer wings of the
galaxy profiles. Since for bright objects, small residuals variations
are not an issue, we simply re-applied our modeling algorithm to the
global background-subtracted images. For bright spirals, the main
issue was that the \wvdecomp\ algorithm splits the galaxy into many
individual objects, corresponding to the surface brightness clumps in
the spiral arms. We visually identified such cases and re-measured
fluxes in elliptical apertures using global background-subtracted
images (see an example shown in Figure~\ref{fig:spiral}). Such cases
are easily identifiable by visual inspection of Magellan images with
overlayed \wvdecomp\ detections. Typically, there are $\sim 10$ such
objects in one image, or $<5\%$ of the total number of galaxies
ultimately used for determination of the cluster stellar mass.

\begin{figure}
\centerline{\includegraphics[width=0.8\linewidth]{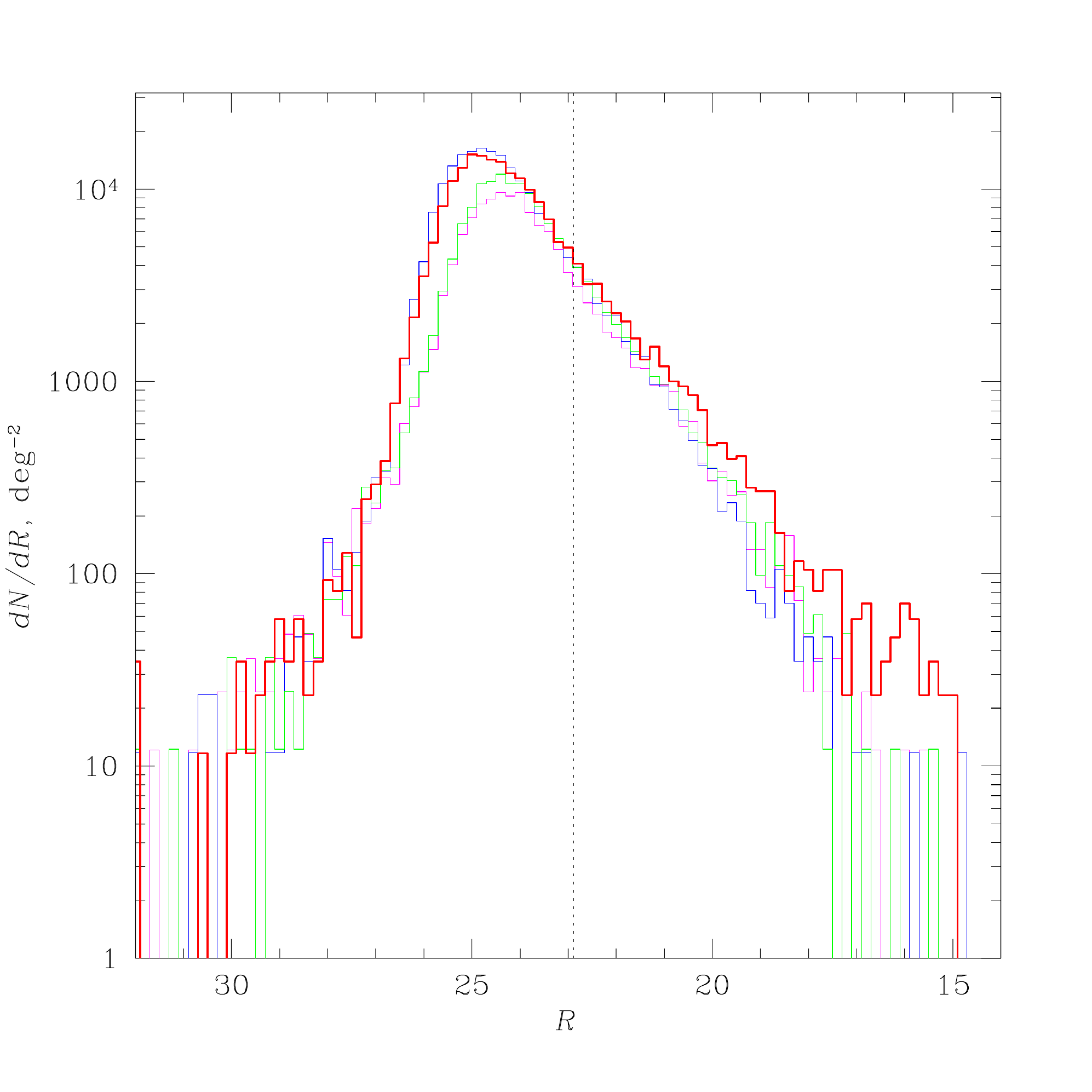}}
\caption{The differential $\log N-\log S$ distribution of detected
  galaxies in the central cluster field (red), and three background
  fields (blue, magenta, and green). The maximum in the $\log N - \log
  S$ histogram provides an estimate for a sensitivity limit in each
  field. The differences in the sensitivity limits primarily reflects
  differences in exposure. The vertical dotted line shows our adopted
  magnitude limit $R=22.9$, which is conservative and applicable for
  all cluster and background fields.}
\label{fig:logN-logS}
\end{figure}

\begin{figure}
\centerline{\includegraphics[width=0.8\linewidth]{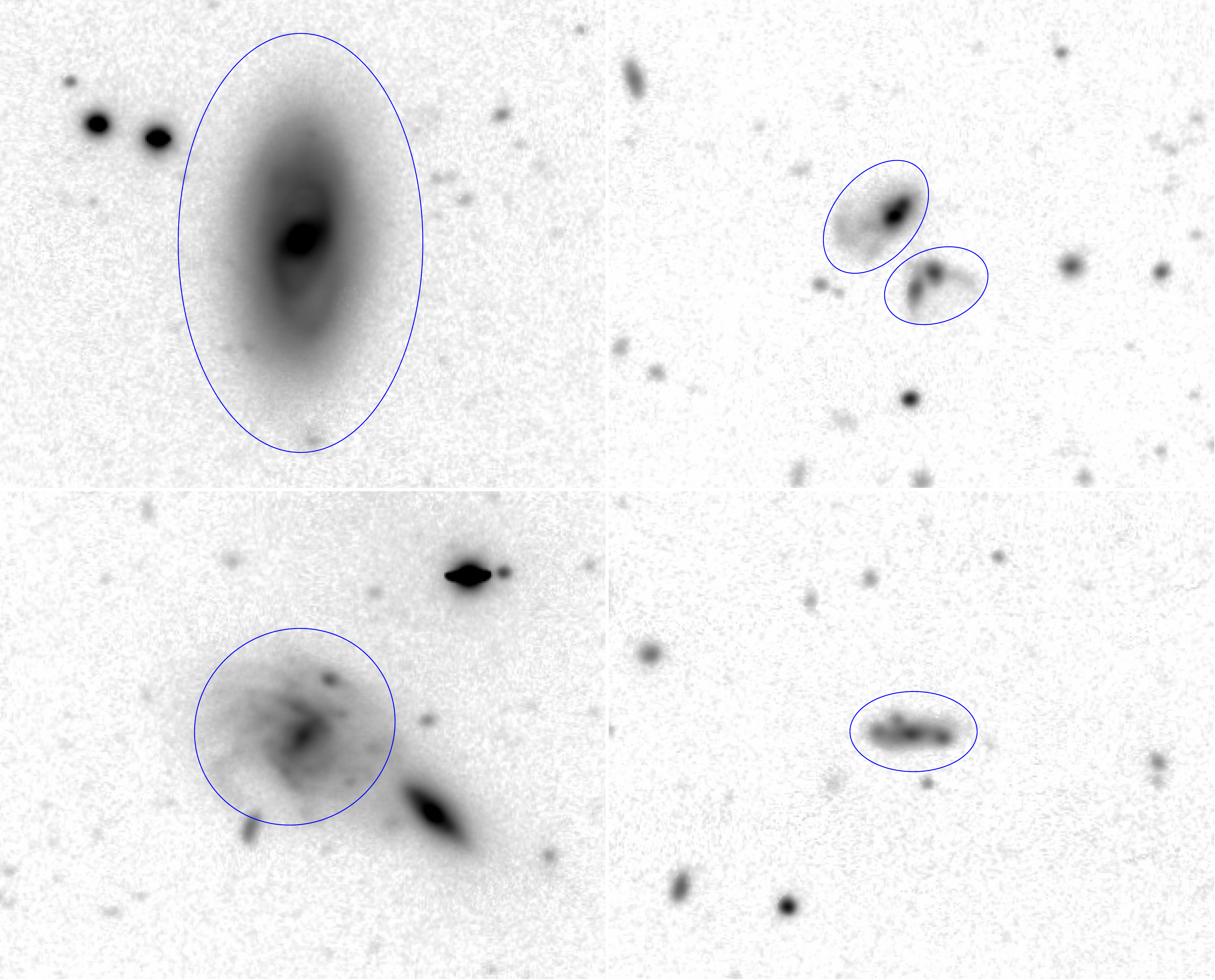}}
\caption{Examples of elliptical apertures used for flux measurements
  in bright spiral galaxies with complex morphology. Our source
  detection algorithm fails for such galaxies because it splits them
  into several objects. Therefore, their fluxes were
  re-measured in hand-set elliptical apertures.}
\label{fig:spiral}
\end{figure}

\subsubsection{Completeness limits}
\label{sec:completness-limits}

The difference in total accumulated exposure in different locations
leads to the differences in the completeness limit. To simplify the
joint analysis of the entire A133 dataset, we need to
define a single completeness limit. We identify completeness for each
field, using the peak location in the differential $\log N-\log S$
distributions. Examples are shown in Figure~\ref{fig:logN-logS}. The
red histogram shows the source counts in the central cluster field,
and the other three histograms show example source counts in the
background fields. The maxima in the $\log N - \log S$ distributions
are well defined, but broad, possibly because of substantial flux
measurement uncertainties near the threshold sensitivity. To avoid
this problem, we set a threshold for further analysis at $\approx 1$
magnitude brighter than the maximal point in the $\log N - \log S$
curves at $m\sim24$. The adopted magnitude limits are $R=22.9$ and
$V=22.6$. In the $R$-band, this limit corresponds to an absolute
magnitude of $M_{R} = -14.1$ at the cluster distance. This is $\approx
5.4$ magnitudes below $M^{*}$ of the $R$-band field galaxy
luminosity function \citep{1996ApJ...464...60L}.

\subsection{Red sequence} 
\label{subsec:Red sequence}

Spectroscopic redshifts are unavailable for the majority of galaxies
detected in the IMACS fields \citep{2018ApJ...867...25C}. Therefore,
we need to subtract the statistical background, which corresponds to
the contribution of foreground and background galaxies to the galaxy
stellar mass functions, cluster light profile, etc. These
contributions were measured in our offset background fields (see
\S~\ref{sec:data} above). The surface number density contrast of the
cluster members relative to the statistical background of galaxies
with the same apparent magnitude is low, except for the central
pointing (e.g., Figure~\ref{fig:logN-logS}). Therefore, we
conservatively use additional selection criteria to remove galaxies
unlikely to be associated with A133.

\begin{figure}
\centerline{\includegraphics[width=0.8\linewidth]{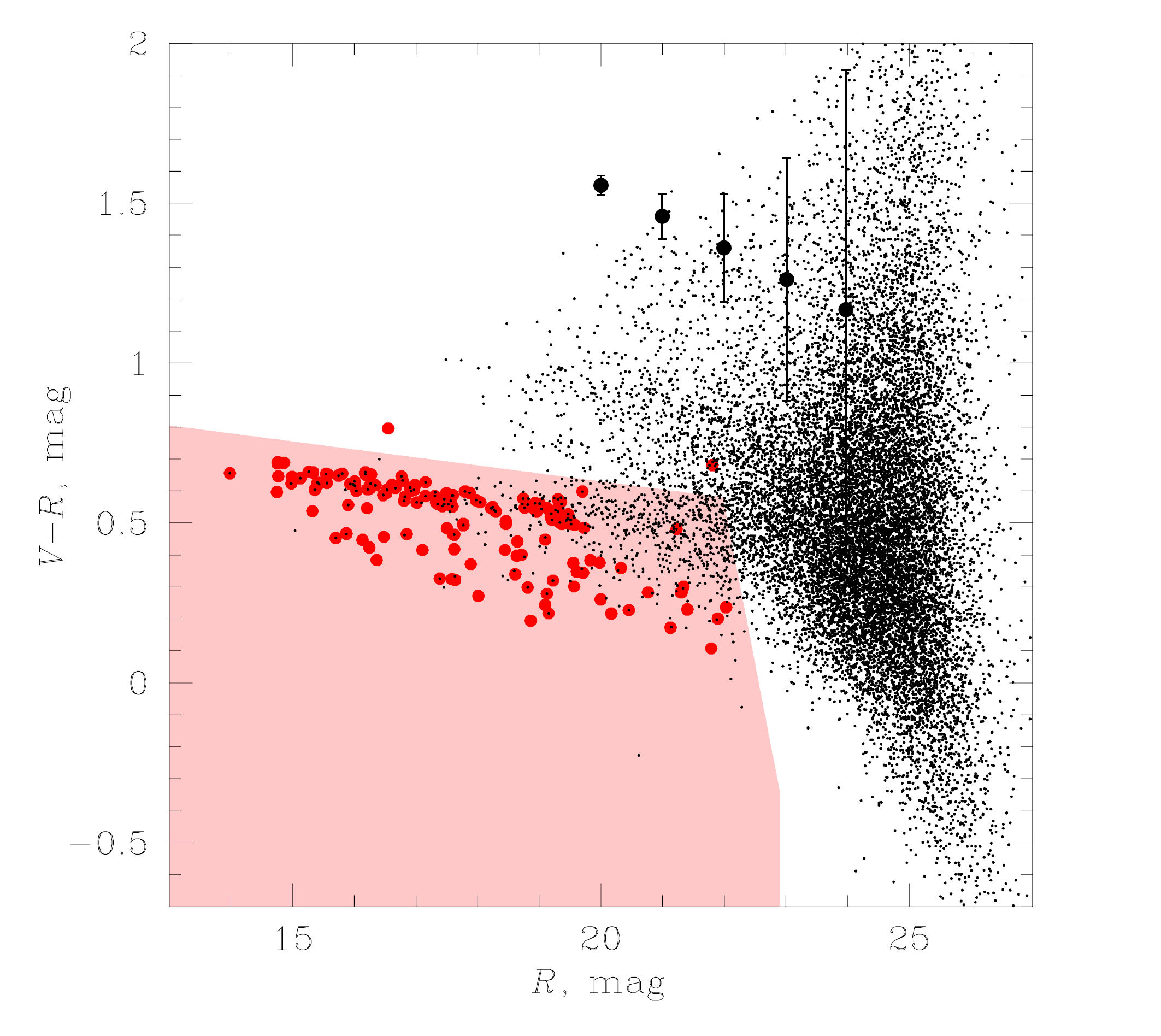}}
\caption{Color-magnitude diagram for galaxies in the central cluster
  field. The shaded region shows our selection criteria for A133
  members. This region is limited by the $R$ and $V$ magnitude limits
  on the right, and by the color-magnitude cut on the top. Red points
  represent cluster galaxies with known spectroscopic redshifts over
  all cluster fields \citep{2018ApJ...867...25C}. The points with
  vertical errorbars indicate typical uncertainties of the V-R color.}
\label{fig:color-mag}
\end{figure}

Our selection is based on the cluster red-sequence method extensively
used for studies and identification of galaxy clusters
\citep[e.g.,][]{1992MNRAS.254..601B, 1992MNRAS.254..589B,
  1998ApJ...501..571G, 2004ApJ...614..679L, 2007ApJ...660..221K,
  2011A&A...536A..34V, 2014ApJ...785..104R}. The underlying idea is
that red, passively evolving cluster members have similar colors and
form a narrow sequence in the color-magnitude diagram. Some of the
galaxies within the cluster, e.g.\ those with recent bursts of star
formation, can be bluer than the red sequence
\citep[e.g.,][]{2019ApJ...875...16C}. However, there should be very
few, if any, galaxies redder than the red sequence members because
those members have the oldest stellar populations. The small number of
objects above the red sequence are ``special cases'', such as
dust-covered AGNs, in which the stellar mass measurements based on
optical luminosities are problematic. Empirically, these
considerations are confirmed by the colors of the A133 members with
spectroscopic redshifts (Figure~\ref{fig:color-mag}).

For the A133 analysis, we used the $V-R$ color-magnitude diagram. This
diagram for the central field (Figure~\ref{fig:color-mag}) clearly
shows the red sequence corresponding to the A133 redshift. We selected
potential cluster members as the objects below or just above the
cluster red sequence, $V-R<1.13-0.025R$, and brighter than the
completeness limits for the $R$- and $V$-band images (recall that
these are $R=22.9$ and $V=22.6$, see
\S~\ref{sec:completness-limits}). These criteria select a pink-shaded
region in the color-magnitude diagram.

We emphasize again that these selection criteria are conservative. To
improve the cluster contrast still further, we could have used a
narrow color band around the red sequence, or additional selection
criteria such as galaxy apparent sizes. However, we then would risk
missing some of the cluster members with unusual properties. Since our
main goal is the maximally complete census of the cluster stars, we
use more inclusive selection criteria described above.

In addition to A133's red sequence, the color-magnitude diagram
seems to show a second sequence with redder colors, $V-R\approx 0.95$
at $R\approx20$. We attribute it to the background cluster projected
onto the core A133, for the following reasons. There is indeed a group
of redder galaxies with smaller diameter in this region
(Figure~\ref{fig:filament}). The brightest of these redder galaxies, at
RA=01:02:45.2 and Dec=$-$21:54:14, has a measured spectroscopic
redshift, $z=0.293$. The location of that second red sequence is in
excellent agreement with the red sequence of Z3146 ($z=0.2906$)
measured by \cite{2007A&A...471...31K} in the same filters. Most of
galaxies in this background cluster should be excluded by our
color-magnitude selection of potential A133 members.

Selection of potential cluster members by means of the color-magnitude
diagram concludes our preliminary data analysis steps. In what
follows, we describe how stellar masses of individual galaxies were
estimated from the optical luminosities, and how they were used to
derive the stellar mass functions and the total stellar mass profiles
within the cluster.

\section{Results: Stellar mass function of the cluster members}
\label{sec:mstar-function}

\subsection{Galaxy stellar mass estimates}
\label{sec:galaxy-stellar-mass}

Using the measured $R$-band magnitudes and $V-R$ colors, we estimated
the stellar mass of each galaxy. The method is based on the stellar
populations synthesis models from \cite{2003ApJS..149..289B}.  The
first step is to convert the apparent magnitudes to the absolute
magnitudes in $R$- and $V$- bands. We used the standard relation,
$M=m-25-\log(D_{L}\,{\rm Mpc})-A-K(z)-EC$, where $D_{L}$ is the
luminosity distance, $A$ is the interstellar extinction correction
from \cite{1998ApJ...500..525S}, $K(z)$ is the $K$-correction, and
$EC$ is the evolutionary correction. The $K$-corrections were obtained
from the ``$K$-correction calculator'' \citep{2010MNRAS.405.1409C,
2012MNRAS.419.1727C}\footnote{Available on http://kcor.sai.msu.ru}.
The evolutionary correction $EC$ is adopted from \cite{1997A&AS..122..399P} 
where it is provided for different galaxy types and a set of photometric bands. We used the $EC$ values averaged over galaxy types. We also note that the evolutionary
correction, $EC=-0.08$, is small at the A133 redshift in both our filters. The described procedure is used to estimate $z=0$ absolute magnitude in the $R$ band and a de-redshifted $M_V-M_R$ color for each galaxy.

To obtain the stellar mass from these parameters, we convert the
$R$-band absolute magnitude to the luminosity in Solar units using
$M_{R,\odot}=4.42$ for the absolute $R$-band magnitude of the Sun
\citep{1998gaas.book.....B}. The luminosity is then converted to the
stellar mass. This conversion is a function of the
$M_V-M_R$ color. Using the \citet{2003ApJS..149..289B} fitting
formulae for $\Mstar/L_R$ as a function of $M_B-M_V$ and $M_B-M_R$, we
obtain (in Solar units):
\begin{equation}
\lg(\Mstar/L_R)=-0.528+1.818(M_V-M_R).
\label{eq:ML:vs:V-R}  
\end{equation}
The main uncertainty in the derived stellar masses is due to
assumptions on the initial mass function of the stellar population
which enter the population synthesis models.
Note that the Bell et al.\ calibration used the ``diet Salpeter'' IMF
model (see discussion in \S 2.4 of \citealt{2018AstL...44....8K}
regarding this choice of the IMF model and its effects on the stellar
mass determination). In this work, we will use the ``diet Salpeter''
based masses. To convert to, e.g., the \cite{2003PASP..115..763C} IMF,
our stellar masses should be scaled down by 0.1 dex. 

\subsection{Stellar mass functions of A133 galaxies}
\label{sec:smf}

Before deriving the total cluster stellar mass as a function of
radius, we need to explore how much light or stellar mass we may be
missing below the completeness limits of our data. We address this
question by analyzing the stellar mass functions of A133 galaxies.

\begin{figure}
\centerline{%
\includegraphics[width=0.5\linewidth]{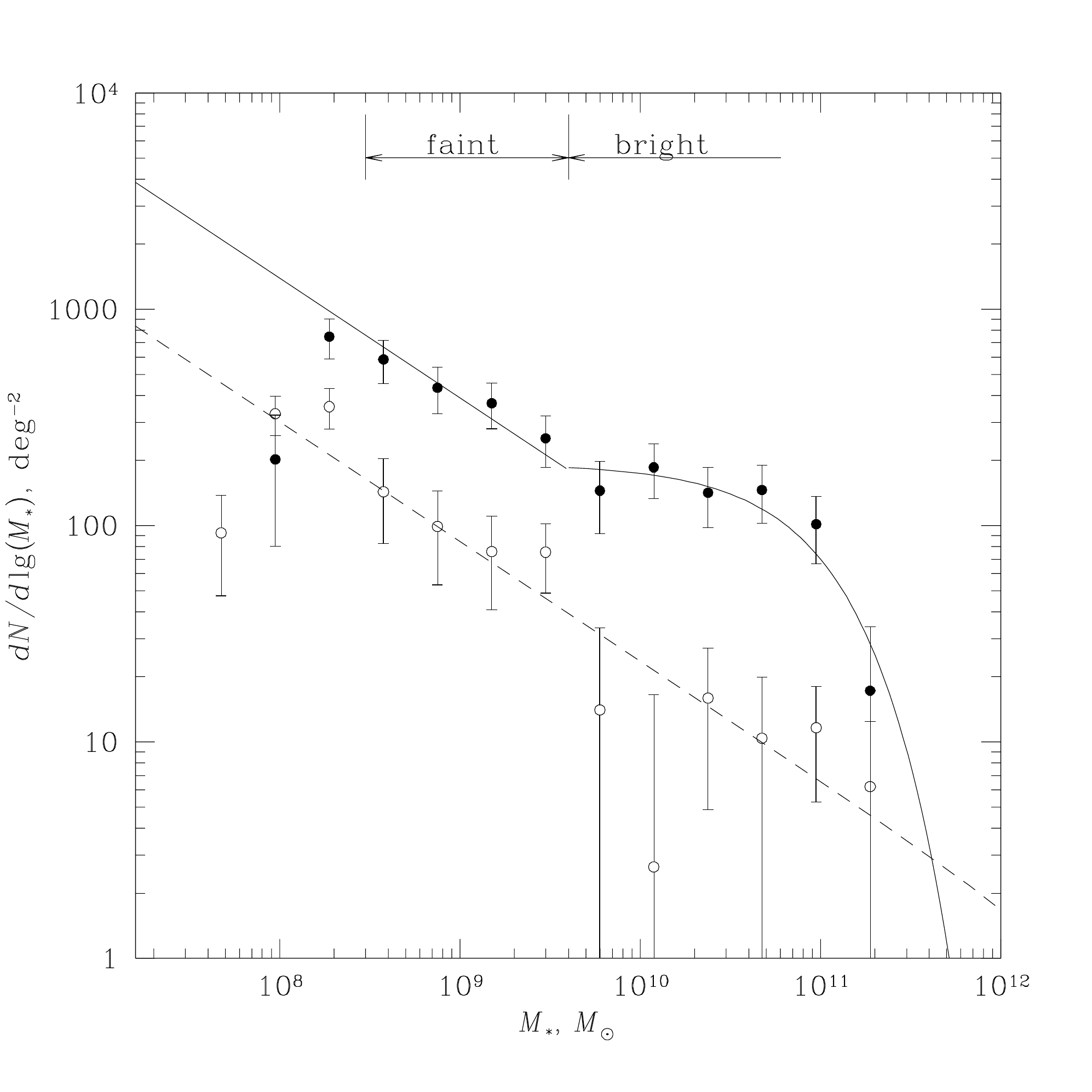}\hfill
\includegraphics[width=0.5\linewidth]{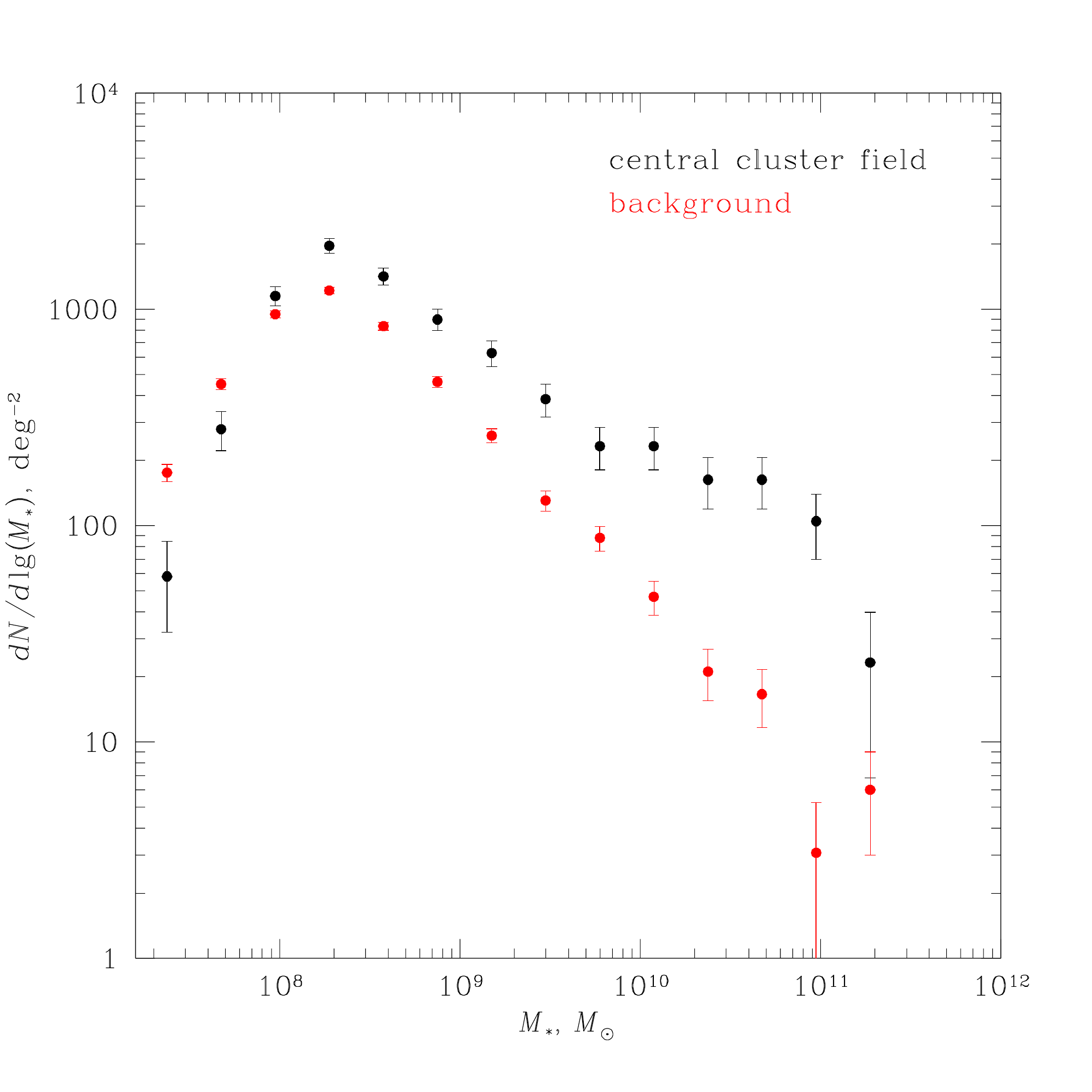}}

\caption{Stellar mass function for galaxies in the central and
  off-center cluster fields (filled and open circles, respectively),
  normalized to the area on the sky. The cluster BCG is excluded from
  this plot. The solid and dashed lines show the best-fit power-law$+$exponential and power-law models (see
  Eq.~\ref{frm:mass_function} and accompanying text). The right panel
  shows the raw data for the central field and the background, to
  illustrate that detection of the two-component structure of the
  stellar mass function is indeed robust. The data for the left panel
  of this figure is available at an online repository \url{https://doi.org/10.5281/zenodo.3610482}.}
\label{fig:mass_function}
\end{figure}

In Figure~\ref{fig:mass_function}, we show the mass functions measured
in the central and off-center cluster fields, normalized to the unit
area on the sky. The background number density of galaxies has been
measured using our off-cluster pointings and subtracted from these
data. The observed mass functions show a roll-over below
$\Mstar\approx 2\times 10^{8} \Msun$. To avoid incompleteness in our
stellar mass measurements, we use only galaxies with
$\Mstar>3\times10^{8}\,\Msun$ in the further analysis. To account for
the stellar mass potentially ``lost'' below this threshold, we
estimate fractional correction factors using analytic fits to the mass
function (see below).

Another prominent signature apparent from
Figure~\ref{fig:mass_function} is a two-component structure of the
mass function in the central pointing (filled circles in the
figure). Above $\approx 4\times10^{9}\,\Msun$, it follows a
Schechter-type \citep{1976ApJ...203..297S} function with an
exponential cutoff around $\Mstar\approx 10^{11}\,\Msun$. At lower
masses the stellar mass function steepens and its behavior is
consistent with a power law down to our completeness limit. However,
the mass function cannot be adequately fit with a single Schechter
model over a broad mass range. The two-component form of the mass
function in the central field is supported by comparison with the mass
function in the off-center fields (open circles in
Figure~\ref{fig:mass_function}). This later mass function is
consistent with a power law with the same slope as that for the faint
end of the mass function near the cluster center. The existence of two
separate components is also supported by the radial profile analysis
in different mass ranges (see \S~\ref{subsec:clust-memb-galax}
below). Following all these considerations, we modeled the central
mass function separately above and below the ``ankle'' at
$\Mstar=4\times10^{9}\,\Msun$. Above this mass, we use a Schechter
model:
\begin{equation}
  \frac{dN}{d\lg M} =  N_0\left(\frac{M}{M_0}\right)^{\alpha}\exp\left(-\frac{M}{M_0}\right)\raisepunct{,}
\label{frm:mass_function}
\end{equation}
where $N_0, M_0, \alpha$ are fitted parameters. In practice, $\alpha$
is unconstrained, given a small dynamical range in masses for the
bright end of the mass function. Therefore, we fixed $\alpha=0$. 

For the mass function in the off-center fields, we used a pure power
law model (dashed line in Figure~\ref{fig:mass_function}). The
best-fit power law slope is $\alpha=-0.55\pm0.11$. The analytic model
of the central mass function in the $3\times10^{8}\,\Msun < \Mstar <
4\times10^{9}\,\Msun$ range is obtained by using a power law with the
same fixed slope $\alpha_{0}=-0.55$ and a normalization chosen to
match the Schechter fit (Eq.~\ref{frm:mass_function}) for brighter
galaxies at $\Mstar=4\times10^{9}\,\Msun$. This fit is also shown by
the solid line in Figure~\ref{fig:mass_function}. It provides a good description to the data. 

We note that the total mass for a power law mass function with a slope
of $\alpha=-0.55$ converges at the faint end. Therefore, there is no
evidence that we may be missing a significant population of dwarf
galaxies which is important for the total stellar mass budget in the
cluster. We also note that if there were a significant population of
undetectable galaxies, especially in the cluster center, it would
contribute to the extended diffuse light envelope, which we analyze
and account for separately (\S~\ref{subsec:bcg}). In the further analysis, we
assume that the power law behavior continues to extremely low masses,
and we apply the corresponding ``incompleteness'' correction of $\times 1.45$ for the
total mass of galaxies with $\Mstar<4\times10^{9}\,M_\odot$. The galaxies with $\Mstar>4\times10^{9}\,M_\odot$ are counted and modeled separately (see below), and they require no incompleteness corrections.

\section{Results: Stellar Mass Profile}
\label{sec:mprof}

Using the results for the stellar mass function of the cluster
members, we proceed to a derivation of the stellar mass profile in the
cluster. Our general approach is to directly count the contribution of
each galaxy with estimated mass $\Mstar>3\times10^{8}\,\Msun$, and
then correct the total for incompleteness at the faint end of the mass
function (see above). We derived the stellar mass profile of the
brightest cluster galaxy separately, tracing the wings of its light
profile to $R\approx 200\,$~kpc. The derived projected profiles of
both the BCG and other cluster galaxies are fit to analytic functions
defined in 3D, and so deprojection is straightforward.

\subsection{Contribution from Cluster Member Galaxies}
\label{subsec:clust-memb-galax}

To determine the stellar mass profile for non-central galaxies, we
computed the surface mass density in radial bins of equal log-width,
centered on the BCG. Since there are indications that the mass
function of the cluster members has two distinct components, we
computed the surface mass density separately in two mass intervals,
$3\times10^{8}<\Mstar<4\times10^9\,\Msun$ and
$\Mstar>4\times10^{9}\,\Msun$. The lower boundary corresponds to the
estimated completeness limit, and the middle point corresponds to the
``ankle'' in the mass function (Figure~\ref{fig:mass_function}).  In
the same mass ranges, we computed the contribution of background
galaxies to the surface mass density, using the data from eight
background fields. This statistical background was subtracted from the
projected cluster mass profiles.

\begin{figure}
\centerline{\includegraphics[width=0.8\linewidth]{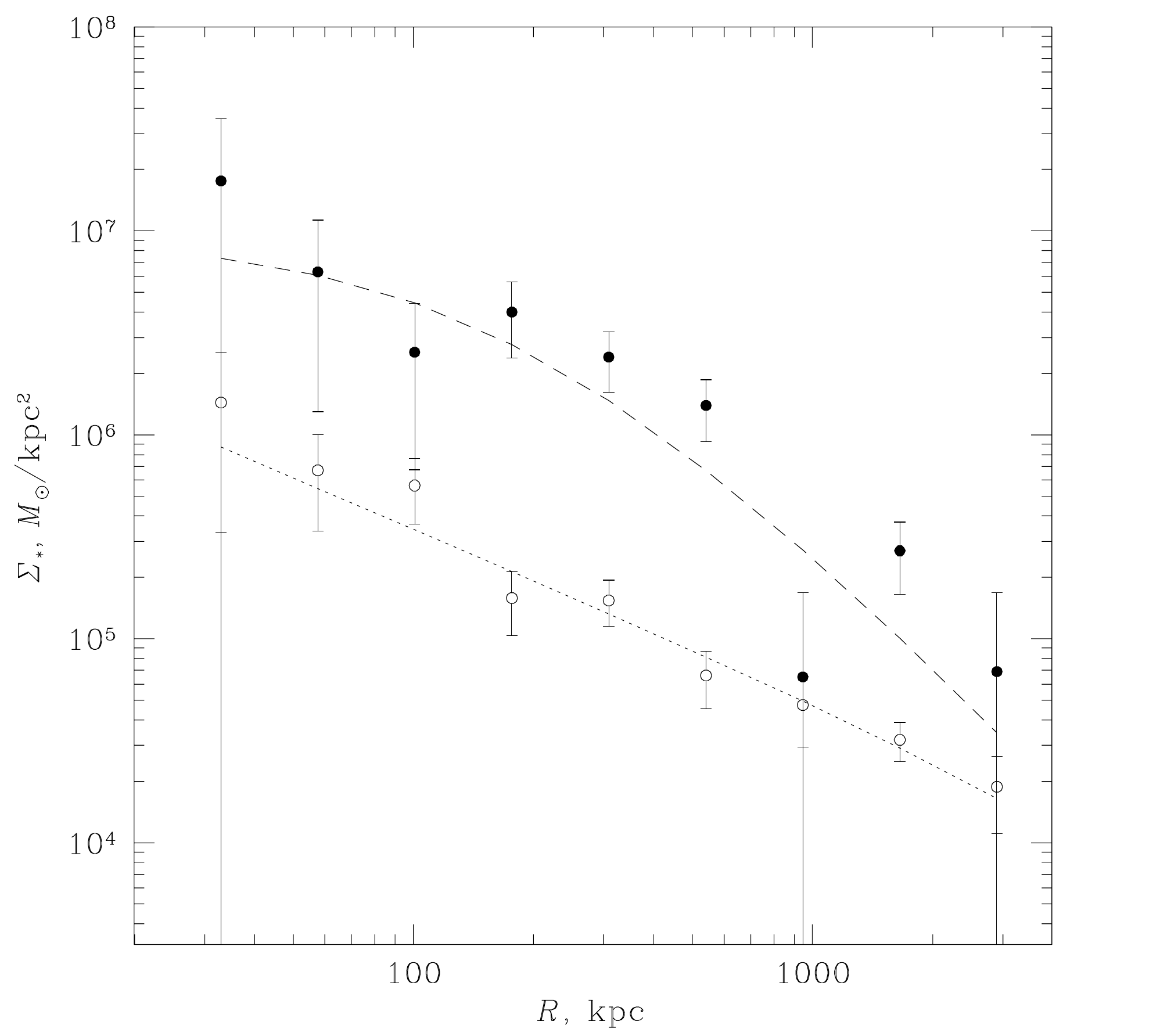}}
\caption{The projected cluster stellar mass profile, derived for
  galaxies with individual masses in the ranges
  $(3-40)\times10^8\,\Msun$ (open circles) and
  $(40-2500)\times10^8\,\Msun$ (filled circles). The lines show the
  best-fit model (see Eq.~\ref{frm:mass_profile} and accompanying
  text). The difference in shape in the two mass profiles is
  insensitive to the exact choice of the boundary between ``faint''
  and ``bright'' galaxies.}
\label{fig:mass_profile}
\end{figure}

The results are shown in Figure~\ref{fig:mass_profile}. Indeed, the
figure shows different radial profiles for the massive and lower-mass
galaxies, which reflects a well known radial dependence of the dwarf-to-giant
galaxy fraction in clusters \citep[e.g.,][]{smith_etal97,2009ApJ...703.2024B}. The lower-mass galaxies have a power-law type profile with a
projected slope of $-0.77\pm0.20$. The profile for massive galaxies
has a flat core out to $R\approx 400-500$\,kpc (nearly
$0.5\,\r500$). Beyond this radius, the profile steepens. While the measured
profiles are quite noisy, there is still a tentative detection of the
cluster signal in both components out to at least $\approx 2$\,Mpc,
beyond the estimated $r_{200c}$ radius.

To deproject these profiles, we fit them using a projected analytic function
defined in 3D. Specifically, we used a ``Nuker'' density profile 
\citep[cf.,][]{1995AJ....110.2622L,1998ApJ...502...48K}, which can be viewed as
a generalized version of the Navarro-Frenk-White \citep[NFW][]{1997ApJ...490..493N} profile: 
\begin{equation}
\rho(x) = \frac{{\rho}_0}{x^{\alpha}(1+x^{\gamma})^{(\beta-\alpha)/\gamma}}\raisepunct{,}
\label{frm:mass_profile}
\end{equation}
where $x=r/{r_s}$, $r_s$ is the scale radius, and $\rho_0$ is the
density scale. The profile inner slope is controlled by $\alpha$,
$\beta$ controls the outer slope, and $\gamma$ determines how fast the
profile slope changes around $r=r_s$. This model has been numerically
integrated along the line of sight and fit to the data.

Since our measured profiles are noisy, we did not fit all parameters
independently. For massive galaxies, we fitted $\rho_{0}$ and $r_{s}$
and fixed the inner slope at $\alpha=0$, the outer slope at $\beta=3$
as expected in the NFW models, and also fixed $\gamma=1$. For the
low-mass galaxies, we used a power-law limit of
Eq.~(\ref{frm:mass_profile}) by fixing $r_{s}$ at a high value. The
best-fit projected profiles are shown as dashed lines in
Figure~\ref{fig:mass_profile}.

The 3D radial profile of the stellar mass in the cluster can be
obtained straightforwardly by integrating the best-fit density profile
of Eq.~\ref{frm:mass_profile}. To estimate the statistical
uncertainties of the derived mass profiles, we used the Monte-Carlo
method described in \cite{2006ApJ...640..691V}. We generated multiple
realizations of the data by scattering the profile data points
according to their statistical uncertainties, re-fit the models,
re-derived the mass profiles, and computed the scatter of mass values
at each radius, averaged over the realizations. The resulting stellar
mass profiles with uncertainties are shown in
Figure~\ref{fig:full_mass_profile} below.

\subsection{Brightest Cluster Galaxy} 
\label{subsec:bcg}

The contribution of the brightest cluster galaxy to the stellar mass
budget was considered separately because this galaxy is bright and
very extended, and therefore cannot be treated as a point mass. We
start by extracting the BCG light profiles in the $R$, $V$, and $I$
filters. Unlike the analysis of non-central galaxies, we used the
global background subtraction (\S~\ref{sec:bg}) because the locally
estimated background subtracts the extended wings of the BCG
profile. Unfortunately, the global background subtraction is less
accurate, leading to increased uncertainties of the BCG profile at
large radii. To estimate the level of these uncertainties, we
extracted $R$, $V$, and $I$-band light profiles around three
representative locations in each of the background and cluster
off-center pointings not contaminated by bright stars.

\begin{figure}
\centerline{\includegraphics[width=0.8\linewidth]{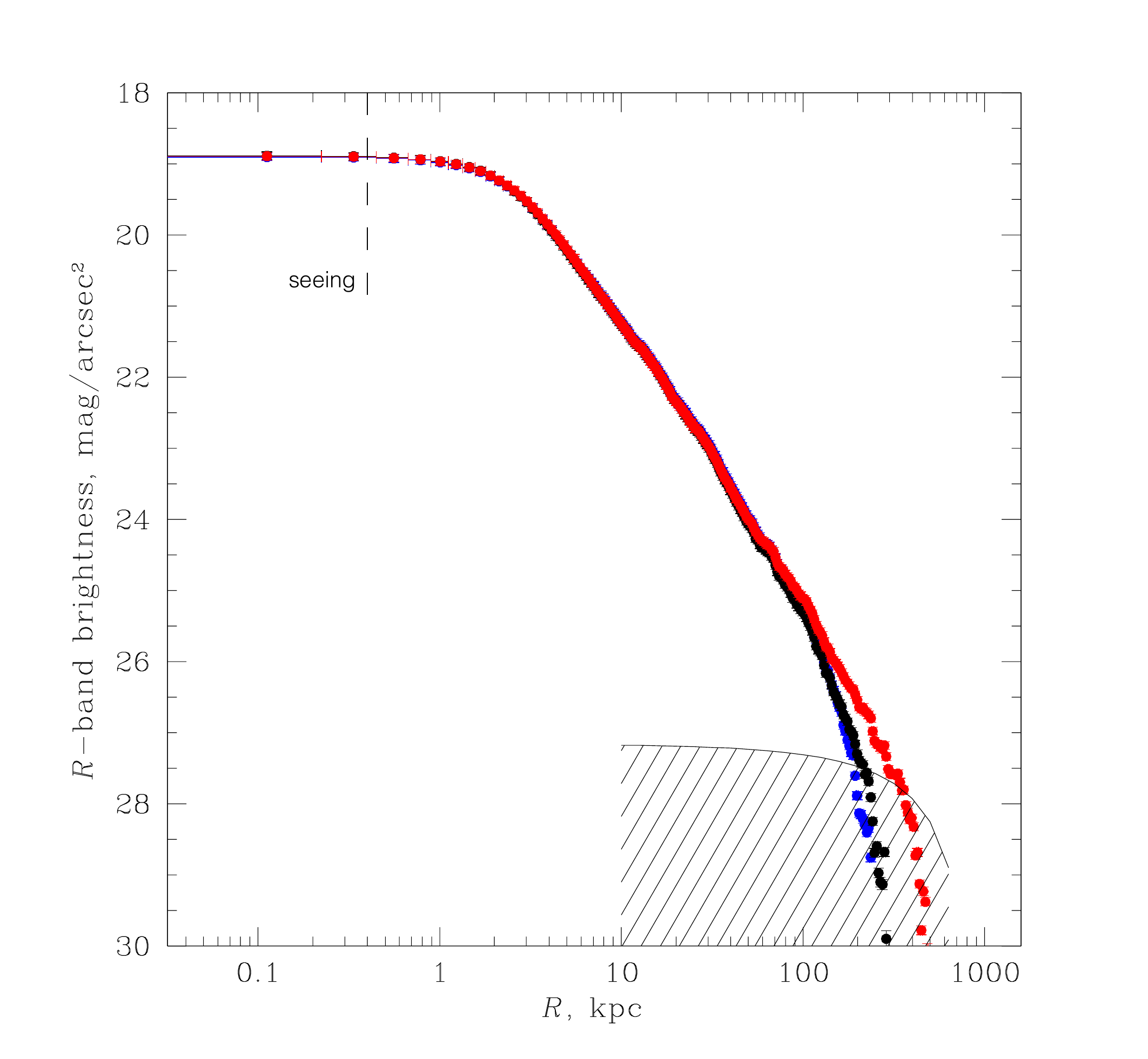}}
\caption{The raw BCG light profile in three filters $R$ (red), $V$
  (blue), and $I$ (black). For clarity, the $V$ and $I$-band profiles
  were renormalized to match the $R$-band profile at small radii. The shaded
  region shows a level of estimated background uncertainties as a
  function of radius (see text).}
\label{fig:light_profile_nocorr}
\end{figure}

\begin{figure}
\centerline{\includegraphics[width=0.8\linewidth]{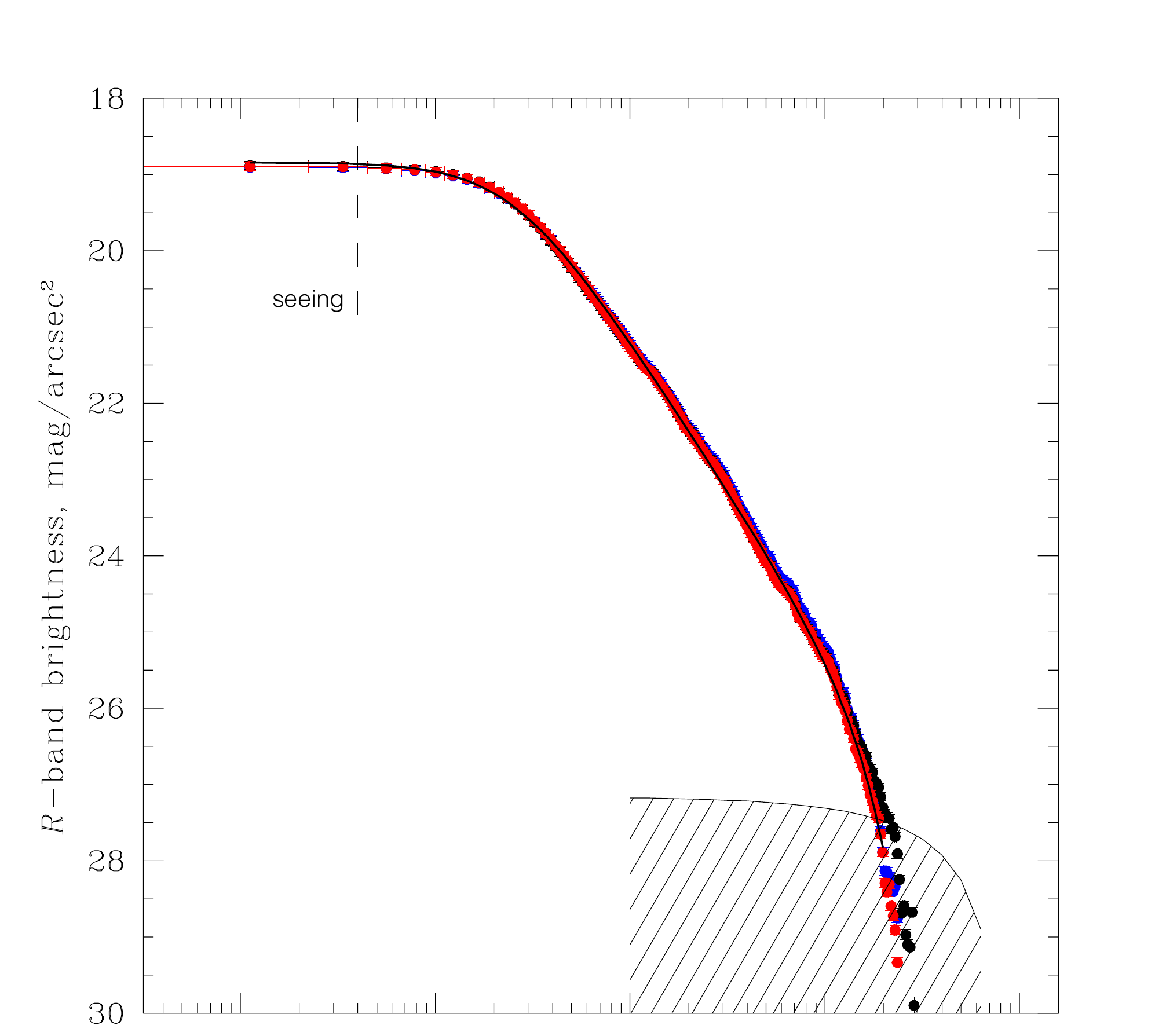}}
\centerline{\includegraphics[width=0.8\linewidth]{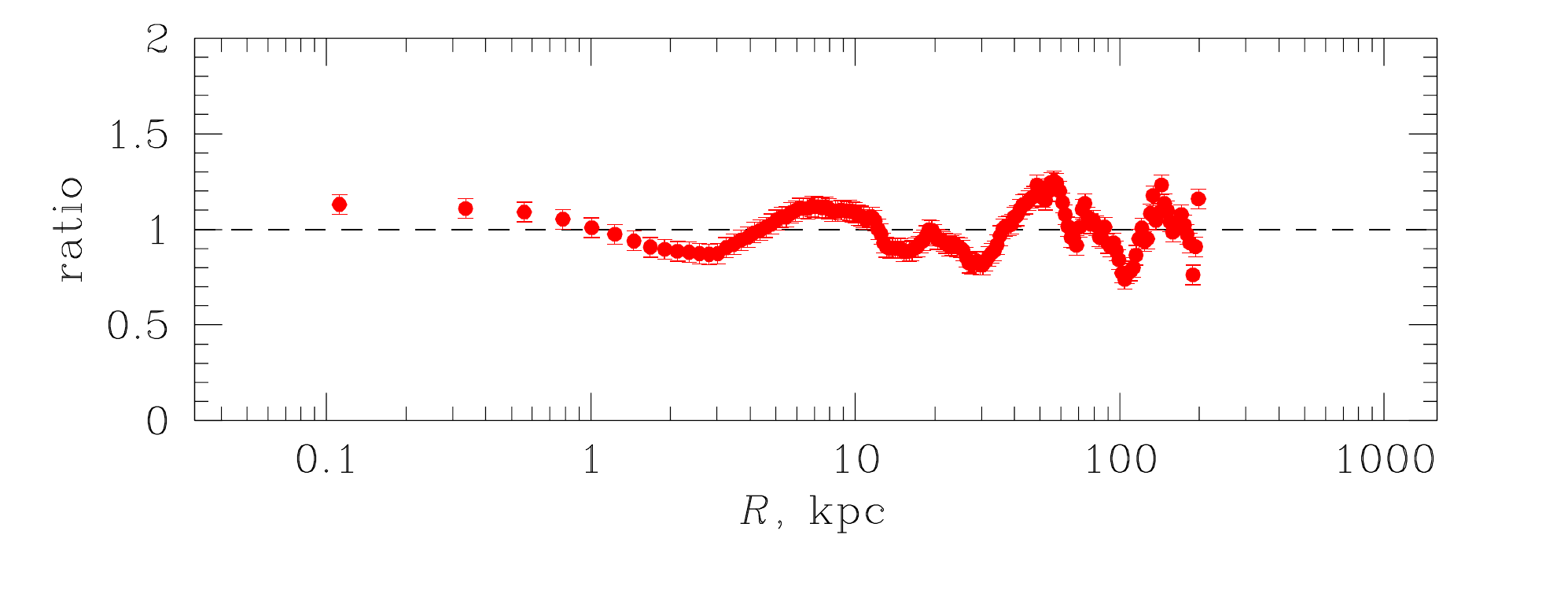}}
\caption{Same as Figure~\ref{fig:light_profile_nocorr}, but with the
  uniform background correction applied to the $R$-band data (see
  text). The black solid line shows the best-fit model (see
  Eq.~\ref{frm:light_profile} and accompanying text). The bottom
  panel shows model residuals.}
\label{fig:light_profile}
\end{figure}

\begin{figure}[tb]
\centerline{\includegraphics[width=0.8\linewidth]{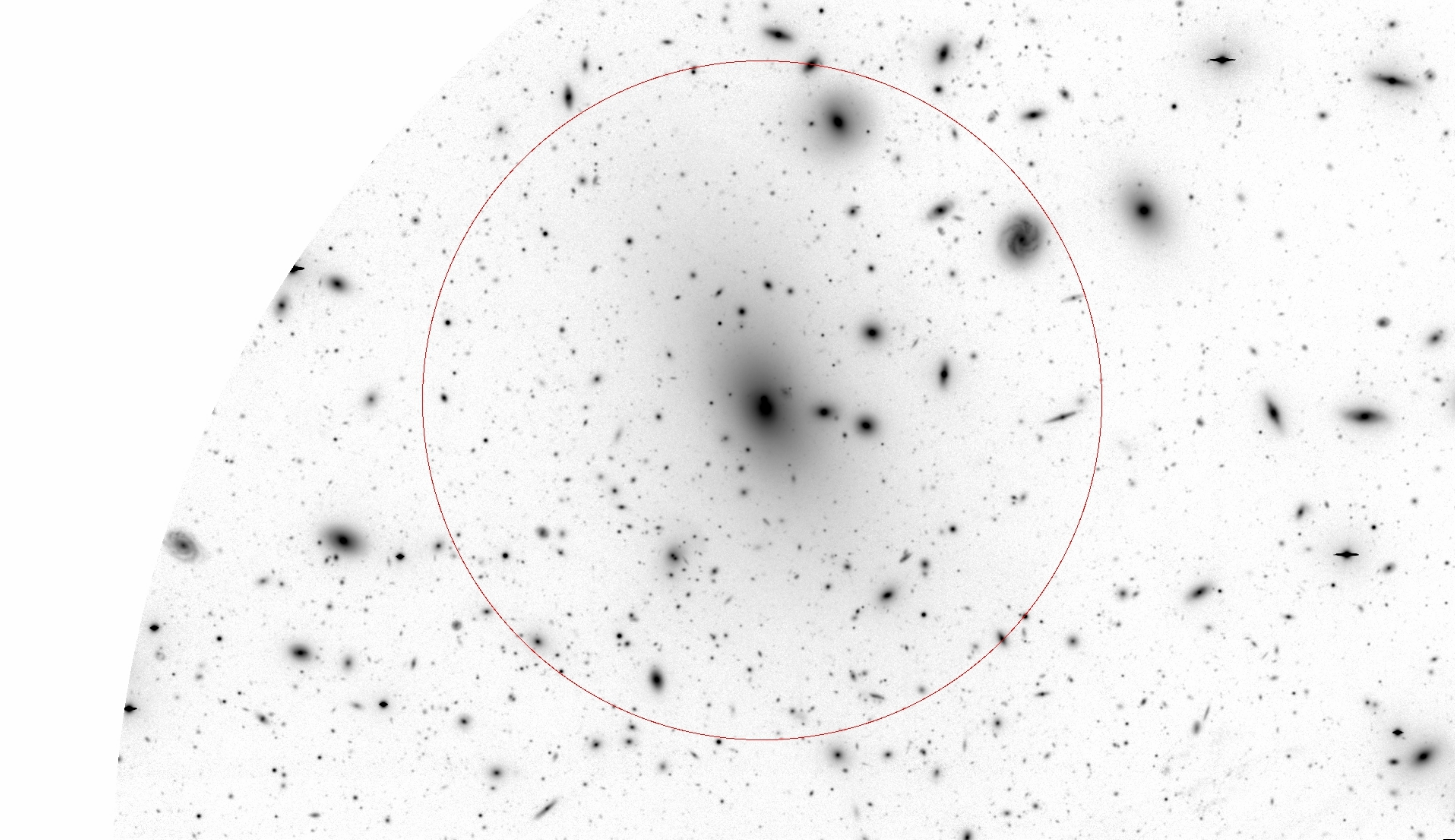}}
\caption{Central cluster region around the BCG galaxy in the
  R-band. Red circle shows a region of $200$~kpc radius where we
  measured the BCG light profile. The wings of the BCG brightness in
  the NE and SW directions can indeed be traced visually very close to
  this radius.}
\label{fig:bcg_image}
\end{figure}

The resulting light profile of the BCG in $R$, $V$, and $I$ filters is
shown in Figure~\ref{fig:light_profile_nocorr}. We have applied
uniform offsets $-0.66$ and $+0.64$ magnitudes per arcsec$^{2}$ to the
$V$ and $I$-band profiles, respectively, to match the $R$-band
brightness at small radii. Note that this BCG $V-R$ color exactly corresponds to the red sequence location for the brightest cluster members (Figure\ref{fig:color-mag}). The level of estimated background
subtraction uncertainties in the $R$-band is shown by a hatched
region. The observed light profiles in all three filters follow one
another very precisely out to $r\approx 150\,$kpc where the $R$-band
brightness reaches a 26~mag~arcsec$^{-2}$ level. Outside this radius,
the $V$ and $I$-band profiles continue to follow one another, while the
$R$-band brightness shows a significant positive deviation. A brightness excess appearing only in the $R$ band is not expected for normal stellar population spectra, where systematic trends run from $V$ through $R$ to $I$. Therefore, a more likely cause of the observed $R$-band excess is inaccuracies in the global background subtraction at
these low surface brightness limits. We find that an additional,
uniform, background correction of 26.9~mag~arcsec$^{-2}$ in the
$R$-band is sufficient to completely match the data in all three
filters (Figure~\ref{fig:light_profile}). This is only 60\% higher
than the typical observed level of background variations at
$R=200\,$kpc, and so such corrections are very likely. We use the
$R$-band profile with this additional background correction in the
analysis below. However, we recognize that the measurements become
extremely sensitive to the background subtraction at $R>200\,$kpc, and
therefore we restrict the mass measurements to within this
radius\footnote{If one assumes that the $R$-band profile should not be
  corrected and integrates it to 400~kpc, this leads to a 33\%
  increase in the estimated BCG stellar mass, or an 11\% increase in
  the stellar mass of A133 within the $\r500$ radius.}. In
Figure~\ref{fig:bcg_image}, we show a zoom-in on the composite
$R$-band image near the BCG location. The 200~kpc radius is shown by
the red circle. The wings of the BCG brightness in the NE and SW
directions can indeed be traced visually very close to this radius.

To reconstruct the 3D stellar mass profile of the BCG, we use an
approach similar to that in \S\ref{subsec:clust-memb-galax}. We fit the
observed light profile with a projected density model defined in 3D and
then integrate that model as a function of radius. In this case, we
used a modified $\beta$-model \citep[c.f.][]{2006ApJ...640..691V}: 
\begin{equation}
\rho(r) = \frac{{\rho}_0}{\left(1+(r/r_c)^2\right)^{3\beta}}\frac{1}{\left(1+(r/r_s)^\gamma\right)^{\epsilon/\gamma}} \raisepunct{,}
\label{frm:light_profile}
\end{equation}
where ${\rho}_0, r_c, r_s, \beta, \gamma, \epsilon$ are fitted
parameters. This model describes a flattening at small radii
($r\lesssim r_{c}$), a transition to a power law behavior at
$r>r_{c}$, and a further change of the profile slope at large radii,
$r\gtrsim r_{s}$. The best-fit model is shown by the solid black line
in Figure~\ref{fig:light_profile} and provides an excellent fit to the
data. Its best-fit parameters are $\rho_0 =
24.1$~mag~arcsec$^{-2}$~kpc$^{-1}$, $r_c = 2.5$~kpc, $r_s =
245.1$~kpc, $\beta = 0.42$, $\gamma = 2.3$, $\epsilon = 0.76$. We use
this model to compute the $R$-band luminosity of the BCG as a function
of 3D radius.

To convert this luminosity to the stellar mass, we use the
mass-to-light ratio from \cite{2003ApJS..149..289B}, just like we did
for the other cluster members. There is only a small change in the
observed color with radius: $V-R=0.66$ at the BCG center, dropping to
$V-R=0.56$ at $r=50\,$kpc, beyond which the contribution of systematic
background uncertainties (see above) makes color gradient measurements
unreliable. Note that such color gradients are quite common in the BCG
and the intracluster light \citep{2018MNRAS.474.3009D}. The
corresponding change in the $M/L$ ratio, from $3.9\,\Msun/L_{\odot}$
in the center to $2.6\,\Msun/L_{\odot}$ at 50\,kpc, was included in
the conversion of the observed light profile to stellar
mass. If, instead, one uses a fixed $V-R$ color measured at
  the center, the BCG stellar mass within 200~kpc is overestimated by 59\%.

Finally, we note that the formal statistical uncertainties of the BCG light
profile within 200~kpc are very small. The mass uncertainties should
be completely dominated by those in the $M/L$ ratio (e.g., those
related to the color gradient or assumptions on the IMF in the the stellar
population synthesis models, c.f.\ \S~\ref{sec:galaxy-stellar-mass}). 

\subsection{Total stellar mass profile}
\label{subsec:total}

\begin{figure}
\centerline{\includegraphics[width=0.8\linewidth]{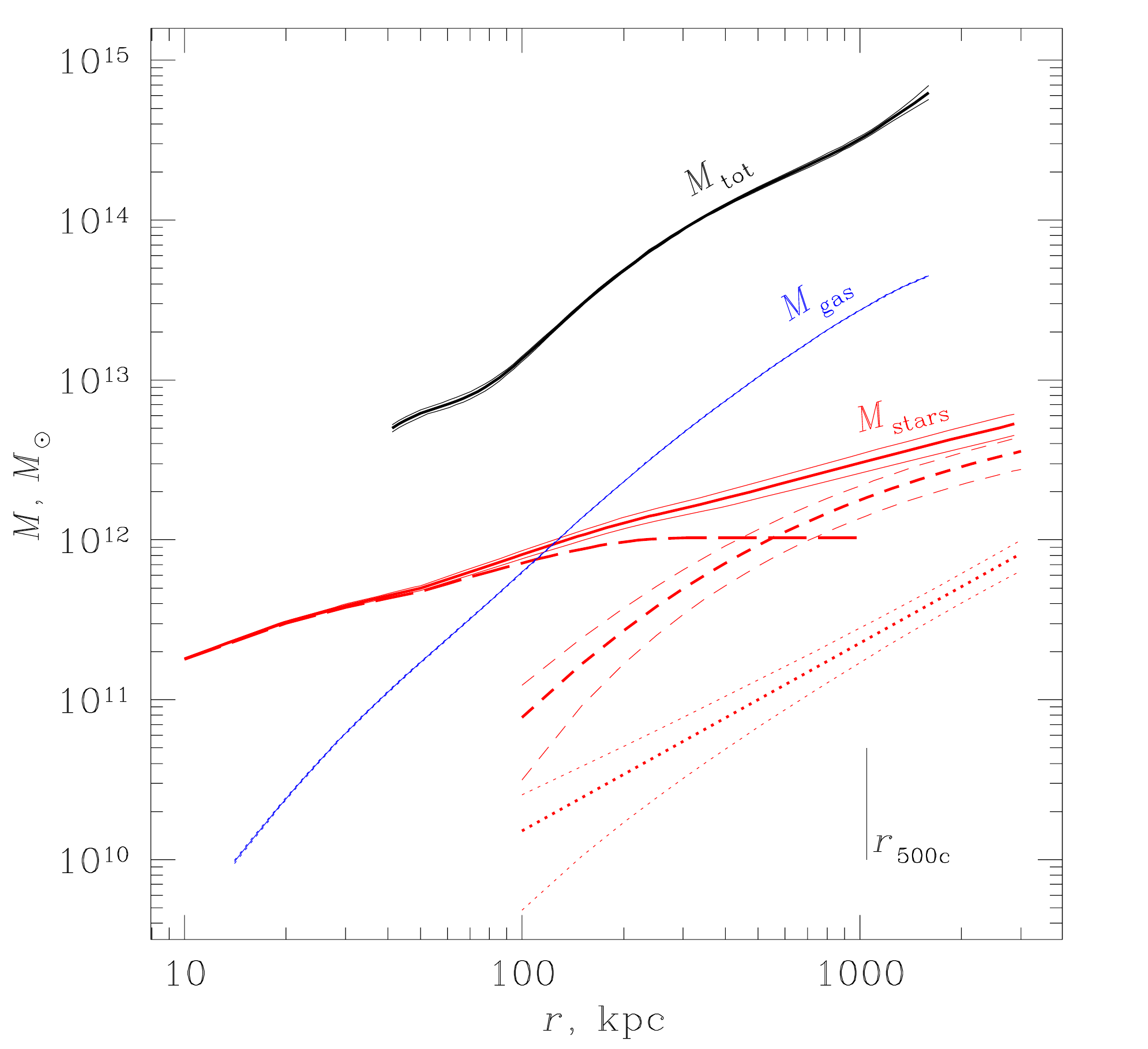}}
\caption{3D profile of enclosed stellar mass of the three stellar
  components: the red dotted line is for low-mass cluster galaxies of
  $M_\star=(3-40)\times10^{8}\,\Msun$, the red dashed line shows high-mass
  cluster galaxies of $M_\star=(40-2500)\times10^{8}\,\Msun$, and the red long
  dashed line represents the BCG. The solid red line shows the summed profile
  of the three stellar components of the cluster. Blue is the X-ray
  estimated mass profile of gas and black line is the total mass
  profile derived from \emph{Chandra} X-ray data using hydrostatic
  equilibrium assumption (Vikhlinin et al.\ in preparation). Short
  vertical black line indicates the $\r500$ radius.}
\label{fig:full_mass_profile}
\end{figure}
In Figure~\ref{fig:full_mass_profile}, we show the total reconstructed
stellar mass profile within A133 and its individual components
discussed above (estimates of $M_{\star}$ within different $r$ are
also presented in Table~\ref{tab:masses}).  For comparison, we also
show the profile of the hot gas and of the total mass reconstructed
from the X-ray data \citep[Vikhlinin et al., in preparation, see
also][]{2006ApJ...640..691V}.  Several points about these profiles are
noteworthy.

The central cluster galaxy contributes a large fraction of the total
stellar mass. Its integrated mass within 200~kpc is approximately 50\%
of the rest of cluster galaxies within 1~Mpc ($\approx \r500$), or
$\sim 25\%$ at $r\approx 3$~Mpc (well outside of $r_{200c}$).  The BCG
dominates the total baryon mass, including hot gas, in the central
$\approx 100$~kpc. The hot gas within this radius shows a spike in
metallicity \citep[see Figure~3 in][]{2005ApJ...628..655V}, likely
reflecting extra enrichment due to stellar mass loss and supernovae
within the BCG.

Non-central galaxies approximately follow the distribution of total mass in the radial range $\approx (0.1-1)\,\r500$ where both the X-ray and optical measurements are most reliable. This is in line with a number of earlier studies for different clusters \citep[e.g.,][]{2015A&A...575A.108A,2016MNRAS.463.1486P}.

Small-mass galaxies of $M_\star<4\times 10^9\, M_\odot$ contribute a
minor fraction of the total stellar mass even at large radii. At
$r=3\,$Mpc, their contribution is $\approx 13\%$ and it is even lower
at smaller radii. The majority of the cluster stellar mass is
contributed by bright galaxies and the BCG.  Nevertheless, the radial
distribution of these faint galaxies is quite distinct from the radial
distribution of brighter galaxies (see
\S~\ref{subsec:clust-memb-galax} above). Namely, the radial
distribution of stellar mass of faint galaxies is well described by a
single power law profile, $\rho(r) \propto r^{-1.77\pm0.20}$. We
discuss possible interpretations of this fact in Section
\ref{sec:discussion} below.

\begin{deluxetable}{p{2.5cm}lllll}
  \tabletypesize{\footnotesize}
  \tablecaption{Stellar mass measurements in Abell 133\label{tab:masses}}
  \tablehead{
      \colhead{$r/\r500$} &
      \colhead{$r$, kpc} &
      \colhead{$M_{\textrm{BCG}}$} &
      \colhead{$M_{\textrm{bright}}$} &
      \colhead{$M_{\textrm{faint}}$} &
      \colhead{$M_{\textrm{stars}}$}
    }
\startdata
0.1\dotfill & 105 & $7.4$ & $0.87\pm0.49$ & $0.16\pm0.11$ & $8.4\pm0.5$ \\ 
0.2\dotfill & 262 & $10.2$ & $4.1\pm1.4$ & $0.47\pm0.21$ & $14.8\pm1.4$ \\
0.5\dotfill & 524 & $10.3$ & $9.7\pm2.5$ & $1.06\pm0.34$ & $21.1\pm2.5$ \\
1.0\dotfill & 1048 & $10.3$ & $18.4\pm4.2$ & $2.4\pm0.6$ & $31.1\pm4.2$ \\ 
1.5\dotfill & 1572 & $10.3$ & $24.6\pm5.5$ & $3.8\pm0.8$ & $38.7\pm5.6$ \\
2.5\dotfill & 2620 & $10.3$ & $33.4\pm7.6$ & $7.0\pm1.5$ & $50.7\pm7.8$
\enddata
\tablecomments{All masses are in units of $10^{11}\,\Msun$ and computed for our
  default cosmology. The last column gives the total of the three
  components --- the BCG, bright galaxies, and faint galaxies.}
\end{deluxetable}

Finally, we note that at $r=\r500$ and beyond, the stellar mass is a
small fraction of the total baryonic mass (i.e., gas + stars),
$\approx 11\%$. This is consistent with the values previously reported
in the literature \citep[e.g.,][]{2013ApJ...778...14G,2018AstL...44....8K}. Therefore,
despite using a much deeper data and an ability to trace the BCG light
profile to larger radii, we have not uncovered a major reservoir of
cluster baryons associated with cluster galaxies. To substantially increase the
fraction of stellar mass in the cluster baryon budget requires drastic
revisions of the $M/L$ ratio values from stellar population
synthesis. Such revisions are not supported by detailed modeling of
the galaxy spectra \citep{2017ApJ...841...68V}. We will present a
detailed analysis of the matter components in A133, including dark
matter and hot intracluster gas, in a subsequent paper.

\section{Discussion of the radial dependence of galaxy stellar mass function}
\label{sec:discussion}

One of the key results of this paper is the upturn of the stellar mass function of satellite galaxies in A133 at $M_\star\lesssim 4\times 10^8\, M_\odot$. The best fit slope $\alpha$ of the power law in this dwarf galaxy regime is  
comparable with recent measurements of the faint-end slope of the stellar mass function
of field galaxies of $-0.5\pm 0.05$  at $z=0.1$ \citep[see, e.g.,][and references therein]{2017MNRAS.470..283W}
and at higher redshifts \citep{2018MNRAS.480.3491W,2018ApJ...854...30P}. 

The existence of the upturn in the luminosity
function in clusters and the value of the faint-end slope 
have been a subject of a long debate in the literature \citep[e.g.,][]{1994MNRAS.268..393D,smith_etal97,2007ApJ...671.1471B,2008AJ....135.1837R,2009AJ....137.3091H,2014MNRAS.444L..34A}, which may be due
to real diversity of the luminosity functions in clusters \citep[e.g.,][]{2015A&A...581A..11M} and, partly, due to degeneracy 
among model parameters in double-Schechter fits to LF. 
Nevertheless, the overall shape of the stellar mass function and its slope in the dwarf galaxy regime are qualitatively consistent with the form of $R$-band luminosity function measured in several nearby clusters \citep{smith_etal97} and groups \citep{zabludoff_mulchaey00}. 
In addition, recent systematic study of luminosity function of galaxies in the SDSS groups and clusters by \citet{2016MNRAS.459.3998L} reported a Schechter$+$power law shape qualitatively similar to that we measured for A133.

Another intriguing result of this study is that the shape of the
stellar mass function changes with radius. This is reflected in the
difference in the radial distribution of low- and high-luminosity
galaxies (Figure~\ref{fig:mass_profile}). It is likely that this
difference is related to the decrease of the ``dwarf-to-giant ratio''
towards cluster center that was previously reported in several clusters
\citep[e.g.,][]{smith_etal97,
  2008ApJ...679L..77S,2009ApJ...703.2024B}.

Cosmological simulations of structure formation in the $\Lambda$CDM
model predict that mass and radial distributions of host halos of
satellite galaxies are nearly self-similar
\citep[e.g.,][]{2004ApJ...609...35K,2005ApJ...618..557N,2016MNRAS.458.2870V,2016MNRAS.457.3492H}
throughout most of cluster volume. Recent study by \citet{han_etal18}
indicates that this self-similarity is broken at $r/R_{200}\lesssim
0.2$, where massive halos have steeper radial distribution resulting
in a smaller dwarf-to-giant halo ratio at these radii. The difference
is due to dynamical friction experienced by massive halos, which
brings them closer to the cluster center.

However, the difference in the radial distribution of dwarf and
luminous galaxies in A133 persists out to $r\sim R_{200}$ and thus
unlikely to be due solely to dynamical friction. The significant
difference in radial and stellar mass distribution of galaxies of
different mass is likely to be yet another manifestation of the break
of self-similarity of galaxy properties due to star formation and
feedback processes accompanying galaxy formation \citep[see][for
reviews]{2015ARA&A..53...51S,2017ARA&A..55...59N}.

One of the potential consequences of feedback in dwarf galaxies is flattening of their central dark matter density profiles \citep{1996MNRAS.283L..72N,2008Sci...319..174M,2012MNRAS.421.3464P} -- the effect that is most efficient for galaxies of stellar mass $M_\star\sim 10^{9}-10^{10}\,M_\odot$ \citep[e.g., see Section 3.1.1 in][for a review]{2017ARA&A..55..343B}. Another effect of galaxy formation that breaks self-similarity is that gas mass fractions are, on average, much larger in dwarf galaxies compared to the giant galaxies. Gas rich dwarfs suffer both tidal stripping and ram pressure stripping of halo and interstellar gas. The latter, if sufficiently fast, can lead to rapid decrease of the gravitational potential depth in the inner regions and significant enhancement of tidal stripping of the stellar component in dwarf galaxies relative to massive ones \citep{2017ApJ...850...99S}.

Interestingly, the radial profile of dwarf galaxies we measured in
A133 can be described by a power law with a slope close to that
expected for the distribution of objects on their first
infall. Indeed, the spherical infall model predicts that before shell
crossing the density profile of matter is $\rho\propto r^{-\gamma}$
with $\gamma\approx 1.5$, while we derive $\rho_{\mathrm{3D}}\propto
r^{-1.77\pm0.2}$ for faint galaxies (\S\ref{subsec:clust-memb-galax}).

Notably, a similar power law radial distribution of blue galaxies was measured in the SDSS \citep{2017ApJ...841...18B} and the DES clusters \citep{2018arXiv181106081S}. Given the expectation for the power law profile of infalling population of matter and galaxies, the most straightforward interpretation of this result is that galaxies do not remain blue much beyond the first pericenter passage and that the population of blue galaxies is thus dominated by infalling galaxies on their first approach to the pericenter, as was previously suggested for dwarf galaxy populations in Fornax \citep{2001ApJ...548L.139D} and Virgo \citep{2001ApJ...559..791C} clusters. Physically, this can happen if star formation of galaxies is decreased and their color reddens on the time scale comparable to the cluster crossing time. The power law distribution of dwarf galaxies we find in A133 can have a similar origin, at least partly,  although the reasons for the disappearance of dwarf galaxies from the sample after the pericenter passage may be different. 

Galaxies can suffer significant morphological transformations and mass
loss due to tidal forces \citep[e.g.,][]{1999MNRAS.304..465M} that
peak strongly around the orbital pericenter. For a given orbit and
strength of the tidal force, stellar systems embedded in a halo with
flattened central dark matter density profile would experience
stronger mass loss and can experience significant increase in the
half-mass radius of the stellar distribution
\citep{2015MNRAS.449L..46E,2017MNRAS.465L..59E}. The latter will lead
to a significant decrease of the galaxy stellar surface density and
surface brightness, potentially bringing it below detection limit of
our observations. Thus, feedback that is expected to flatten dark
matter distribution predominantly in the centers of dwarf galaxies of
$M_\star\sim 10^{9}-10^{10}\,M_\odot$ may affect dwarf and luminous
galaxies very differently, thereby breaking the self-similarity of
gravitational collapse. Indeed, \citet{2011MNRAS.416.1197W} compare
results of the semi-analytic models used with cosmological simulations
of clusters that match observed dwarf-to-giant galaxy ratios in Virgo,
Fornax, Coma, and Perseus clusters and conclude that the tidal disruption
of dwarf galaxies needs to be enhanced in the models. Observations also
show indications that low surface brightness galaxies suffer
significant tidal stripping and disruption in the central regions of
clusters \citep[e.g.,][]{2017MNRAS.470.1512W}.

Another interesting fact is that in group-scale halos the
dwarf-to-giant ratio appears to be enhanced compared to the field
\citep{zabludoff_mulchaey00} -- the trend opposite to that found in
massive clusters, and which is reflected in the systematic change of
the shape of the luminosity function from rich clusters to groups
\citep{2016MNRAS.459.3998L}. This trend can be understood as a net
result of two opposing trends: the increased efficiency of tidal
disruption of dwarf galaxies in massive clusters due to stronger tides
and larger rate of disruption of massive galaxies in groups due to
more efficient dynamical friction.

\section{Summary and Conclusions}                      
\label{sec:conclusions}

This paper presents the analysis of deep optical imaging observations of the galaxy cluster Abell 133 with Magellan/IMACS. The summary of our main results is as follows:

\begin{itemize}
\item The stellar mass function of cluster member galaxies is reliably
  measured down to a mass limit of $M_\star=3\times10^8\,M_\odot$
  ($\approx 0.1$ of the LMC stellar mass). The mass function shows a
  clear two-component structure with an excess of
  $M_\star<4\times10^9\,M_\odot$ galaxies over an extrapolation of the
  Schechter fit from higher masses. There is a background cluster
  ($z=0.29$) projected on the center of A133, but based on the spatial
  distribution of faint galaxies we confirm that the low-mass
  component is associated with A133 itself.
\item Interestingly, the radial profile of dwarf galaxies
  ($M_\star<4\times10^8\,M_\odot$) we measured in A133 can be
  described by a power law with a slope of $-1.77\pm 0.2$. This is
  close to the power law radial distribution with the slope of
  $\approx -1.5$ expected for objects on their first infall in the
  spherical infall model of cluster formation. This similarity may
  indicate that dwarf galaxies are disrupted efficiently in clusters
  and that most of them do not survive for more than a single
  orbit. However, additional observational measurements and more
  detailed modelling is required to test this conjecture.
    
\item We have measured an extended halo of the brightest cluster
  galaxy to $\sim 200$ kpc. Its profile is fully within the range of
  BCG envelopes measured for other low-z clusters
  \citep[e.g.,][]{2018AstL...44....8K}. Including the outer envelope,
  the BCG contributes $33\%$ of the cluster stellar mass within
  $r_{500c}$, also in the range of previously observed values
  \citep{2000ApJ...536..561G,2018AstL...44....8K,2019ApJ...874..165Z,2019arXiv190808544K,2020MNRAS.491.3751D}.

\item The total stellar mass has been measured in a range of radii out
  to $\approx 2.5\,$Mpc with formal statistical uncertainties of
  $<15\%$. The dominant source of systematic uncertainty is the
  stellar mass-to-light ratio. We have used the $V-R$ color dependence
  of $M/L$ from \cite{2003ApJS..149..289B} computed using population
  synthesis models and corrected to a ``diet Salpeter'' stellar initial
  mass function. For comparison, the $M/L$ values for the
  \cite{2003PASP..115..763C} IMF would already be $\sim 25\%$
  lower. Detailed studies of the impact of the $M/L$ assumptions on
  the stellar mass measurements in A133 are beyond the scope of this
  paper.
\end{itemize}

\acknowledgments

This paper includes data gathered with the 6.5 meter Magellan
Telescopes located at Las Campanas Observatory, Chile, and public
archival data from the Dark Energy Survey (DES).  The authors thank an anonymous referee for the constructive suggestions that helped to improve this paper. We are grateful to Ann Zabludoff for useful discussions related to existing observational
constraints on the dwarf-to-giant galaxy trends in galaxy groups and
clusters. S.\ S.\ and R.\ K.\ were supported by NASA contract
NAS8-03060. A.\ K.\ was supported by the NSF grant AST-1714658 and by the
Kavli Institute for Cosmological Physics at the University of Chicago
through grant PHY-1125897 and an endowment from the Kavli Foundation
and its founder, Fred Kavli.  The work of T.\ C.\ was carried out at
the Jet Propulsion Laboratory, California Institute of Technology,
under a contract with NASA.

\appendix

\section{\wvdecomp\ overview and empirical noise maps}
\label{sec:wvdecomp-and-noise}

\begin{figure}
\centerline{\includegraphics[width=0.99\linewidth]{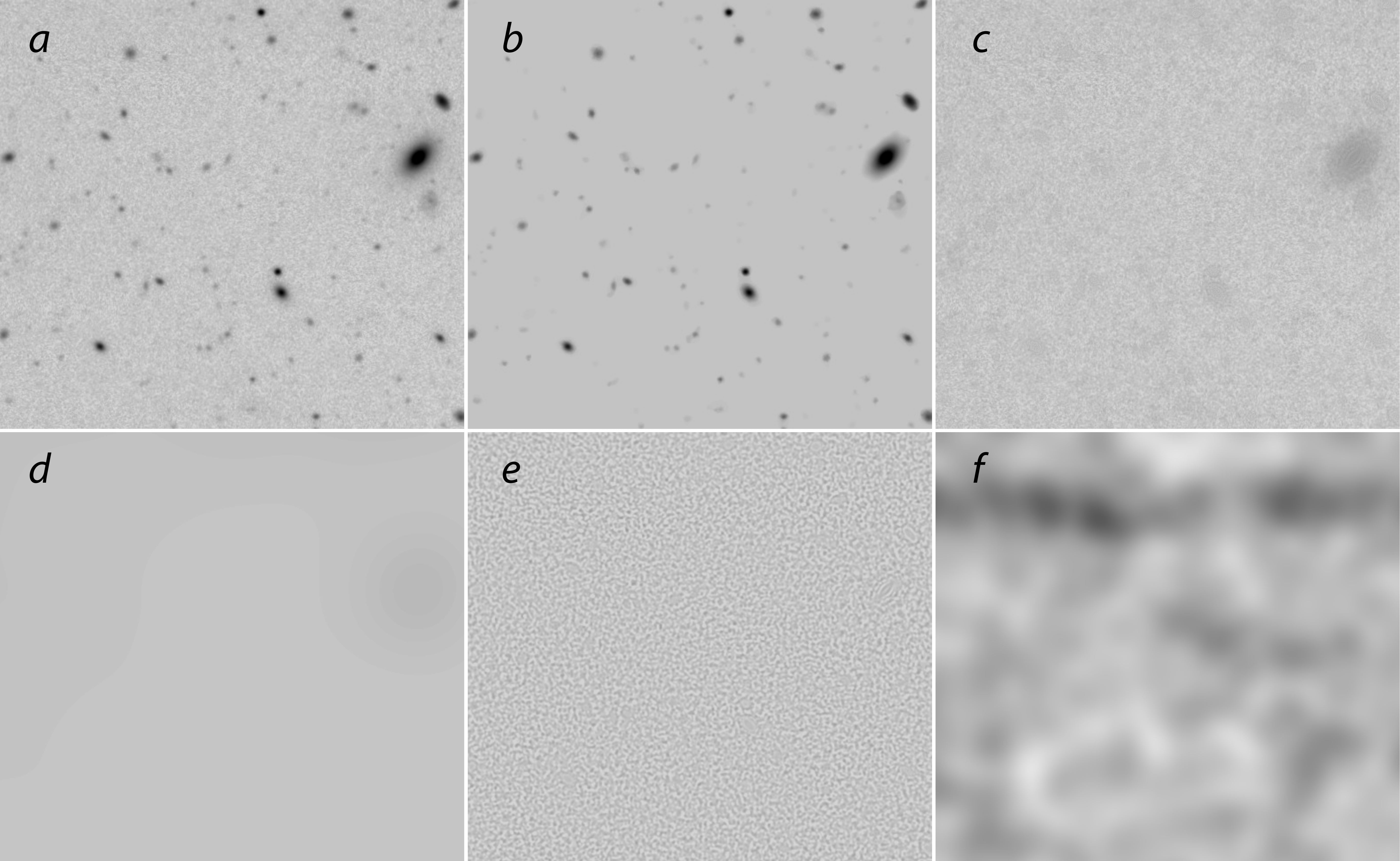}}
\caption{Steps in making the spatially-nouniform noise maps for source
  detection. The flat-fielded, background-subtracted image (panel
  \emph{a}) and a first-iteration noise map (see text) serve as inputs
  to the initial run of \wvdecomp. The output (panel \emph{b}) is
  detected structures on spatial scales $\sim 0.2''-7''$; this map is
  0 outside of the ``islands'' defined by detected structures. Panel
  \emph{c} is the source-cleaned image, obtained as a difference of
  maps in panels \emph{a} and \emph{b}. Panel \emph{d} is the
  convolution of data in panel \emph{c} with a Gaussian of
  $\sigma=25''$; this served as the local background map (see
  \S~\ref{sec:bg}). Panel \emph{e} is the convolution of data in panel
  \emph{c} with the \wvdecomp\ kernel on the $0.4''$ spatial scale. It
  is an estimate of pixel-to-pixel noise in the original image, modulo
  a renormalization factor of 0.20066. Panel \emph{f} is the final
  noise map obtained by smoothing the square of the data in panel
  \emph{e} with a  Gaussian kernel with $\sigma=3''$.}
\label{fig:noise:map}
\end{figure}

\wvdecomp\ is the wavelet-based algorithm for finding stastically
significant structures in astronomical images and separating them into
a range of spatial scales of interest\footnote{\wvdecomp\ is available
  at an online repository archived at DOI: 10.5281/zenodo.3610345 \citep{https://doi.org/10.5281/zenodo.3610345}.}. For full reference, see
\cite{1998ApJ...502..558V}. Here we review the outputs produced by
\wvdecomp\ and explain how these were used to compute
spatially-dependent noise maps (\S~\ref{sec:detection}). 

The procedure is illustrated in Figure~\ref{fig:noise:map}. The first
run of \wvdecomp\ over the input flat-fielded and
background-subtracted image (shown in panel \emph{a}) uses an
approximate noise map, computed from the mean rms deviations over the
full image area and only corrected for exposure variations. One of the
outputs of \wvdecomp\ is the image containing identified statistically
significant structures (in this case, on spatial scales $\sim
0.2''-7''$; see panel \emph{b}). An equivalent image at the end of
this procedure can be used to identify sources simply by finding local
maxima in this \wvdecomp\ output, as shown in
Figure~\ref{fig:detections} above. This image can also be used to
identify ``islands'' of significant signal around each detected
sources. Such islands are useful for selecting image subsections for more
detailed modeling, and for masking out unrelated
sources. Identification of the islands is straightforward for isolated
sources. In case of the overlapping sources, a version of the ``water
fill'' algorithm can be used \citep{waterfill2013}. A similar
algorithm is used for source de-blending in \emph{SExtrator}
\citep{1996A&AS..117..393B}.

Here, we use the \wvdecomp\ output to compute the source-cleaned image
which retains all of the noise (panel \emph{c}). A convolution of this
source-free image with a wide Gaussian (panel \emph{d}) gives an
estimate of the local background (c.f.\ \S~\ref{sec:bg}). Note a
slight enhancement of the estimated background at the position of a
brighter ellptical galaxy near a top-right corner of the image. This
enhancement is insignificant in this case, but becomes a problem for
the brightest galaxies and the cluster BCG, in which cases we used the
global background (\S~\ref{subsec:fluxes and magnitudes}). A
convolution of the same source-free image with the \wvdecomp's wavelet
kernel, which is then squared, appropriately renormalized, and
smoothed with a $\sigma=3''$ Gaussian, provides an estimate of the
spatially-variable noise (panel \emph{f}). This map serves as an input
to the final run of \wvdecomp\ leading to source detections (see
Figure~\ref{fig:detections} for comparison).

\section{Comparison of stellar masses derived from Magellan and DES data}
\label{sec:desVSmag}

\begin{figure}
\centerline{\includegraphics[width=0.8\linewidth]{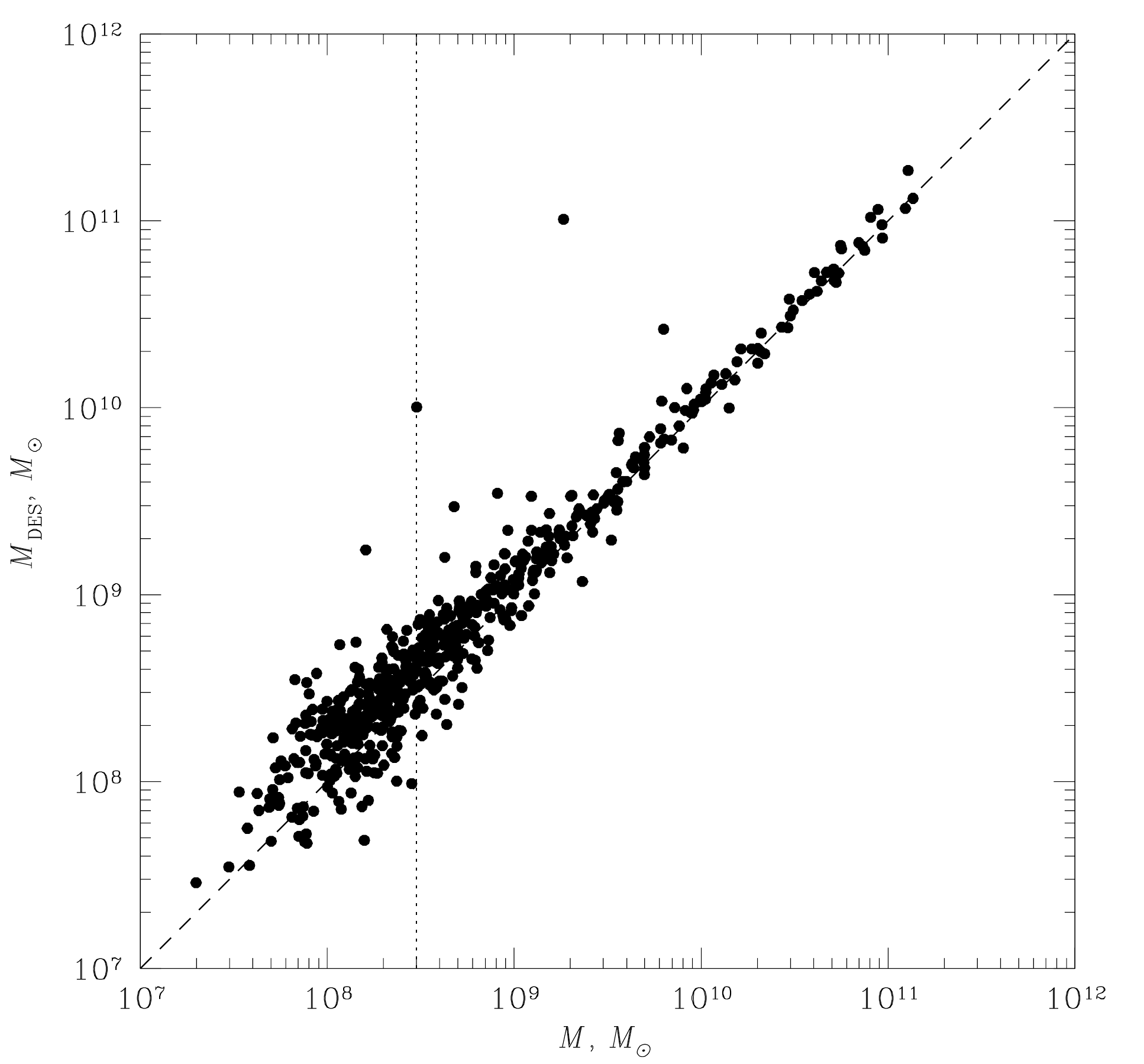}}
\caption{Comparison of galaxy stellar mass measurements from this
  study and those based on the DES photometries. The black dotted line
  indicated our adapted lower mass limit of $3\times10^8\,\Msun$
  (\S~\ref{sec:mstar-function}).}
\label{fig:mass_mass}
\end{figure}

In order to ensure the reliability of our measured galaxy masses we
compared them with masses derived from photometric data of the Dark
Energy Survey (Data release 1) \citep{2018ApJS..239...18A}. We
selected the DES sources around the Abell 133 BCG position and matched
with our catalogue in the central cluster field. Our sources belong to
the red sequence and considered to be cluster galaxies. The DES did
not observe in $V$; therefore we used $g$ filter instead. We converted
DES $g$ and $r$ magnitudes into \emph{sdss} magnitudes, applied the
$K$-correction, transformed magnitudes to luminosities and then to
masses using \cite{2003ApJS..149..289B} expressions. Derived DES
masses and our measured masses from the Magellan data are plotted in
the Figure~\ref{fig:mass_mass}. Outliers with overestimated DES masses
are located near the BCG and a bright galaxy, or parts of double
sources.

\begin{figure}
\centerline{%
\includegraphics[width=0.49\linewidth]{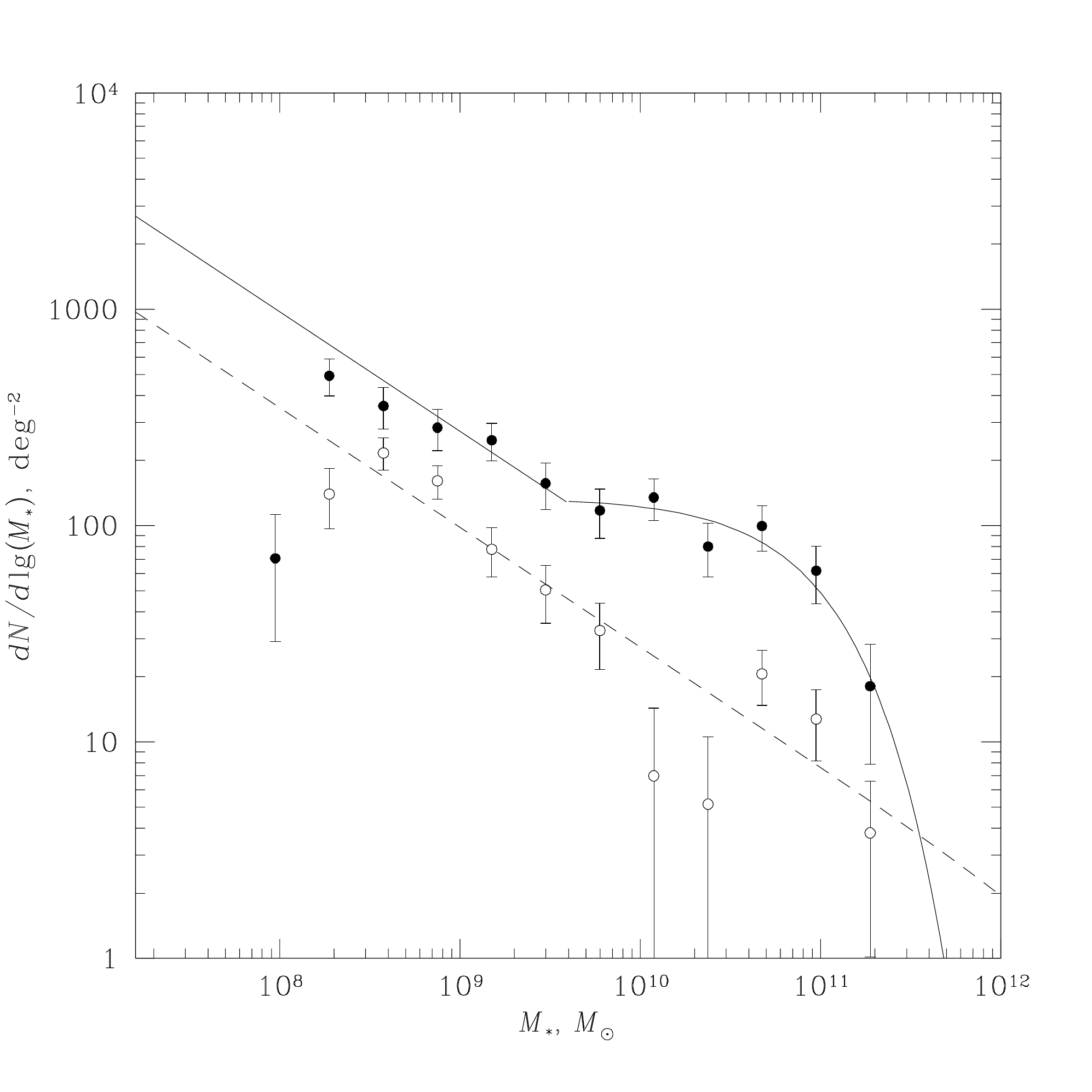}\hfill
\includegraphics[width=0.49\linewidth]{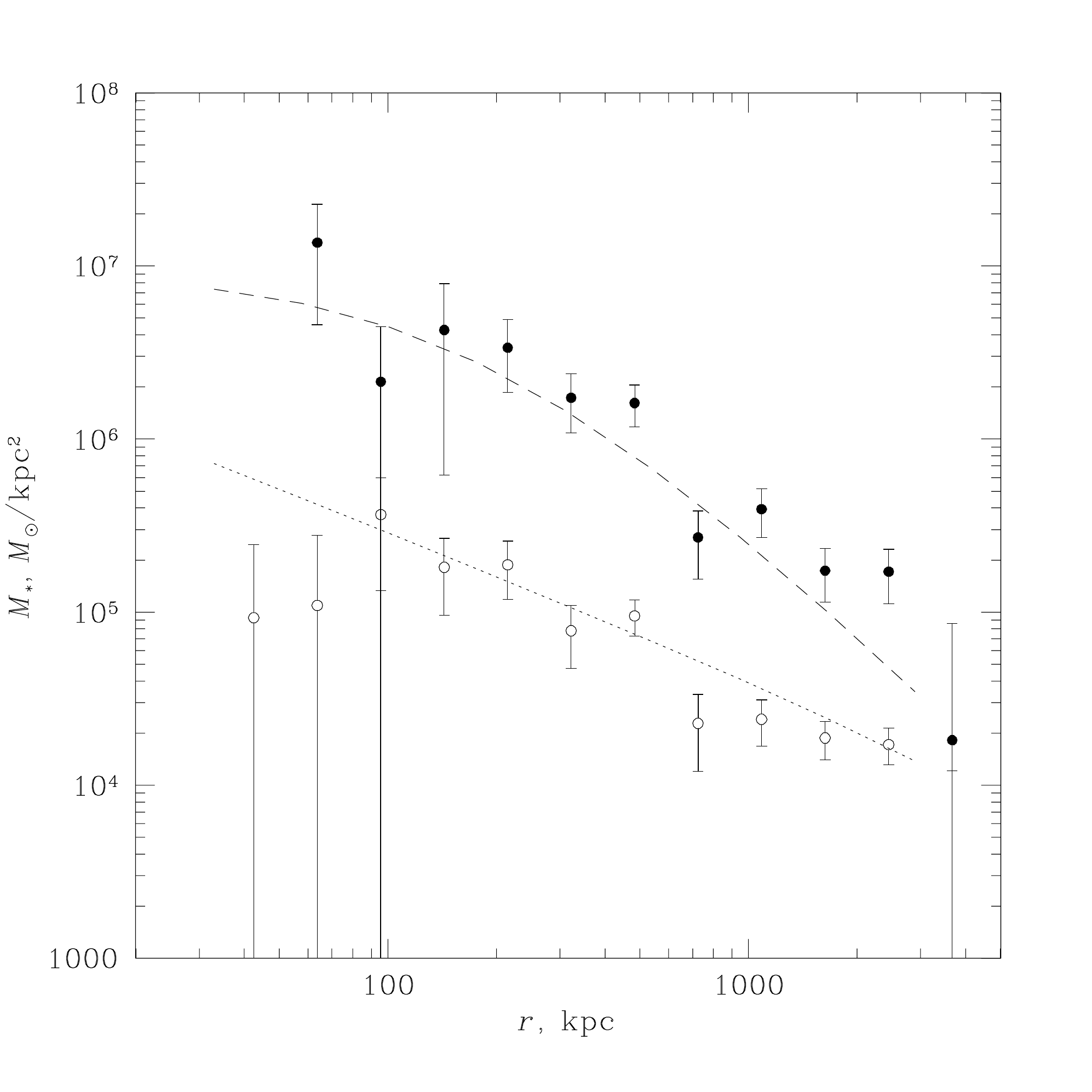}%
}
\caption{Stellar mass functions and the projected mass profiles for
  non-central galaxies derived from the DES data. The figures are
  equivalent to Figure~\ref{fig:mass_function}
  and~\ref{fig:mass_profile}. The lines show analytical fits to our
  Magellan data. Completeness of the mass function is reached at a
  higher  threshold compared to Magellan data, so we used the mass
  range $\Mstar=(6-40)\times10^{8}\,\Msun$ for the ``faint galaxies''
  profile shown in the right panel. A somewhat lower normalization of
  this profile relative to the Magellan fit is fully consistent with
  the difference in the incompleteness corrections (cf.\
  \S\ref{sec:smf}), 1.74 vs.\ 1.45 for the
$\Mstar=(3-40)\times10^{8}\,\Msun$ mass range we used for Magellan data.}
\label{fig:DES_results}
\end{figure}

We have repeated the entire analysis chain presented in this paper
using DES catalogs. The DES data are shallower, but cover the entire
cluster region. The stellar mass functions and the mass profiles
derived from DES are fully consistent with our measurements
(Figure~\ref{fig:DES_results}). We could not use DES data for fitting
the outer envelope of the BCG because of the over-subtraction of the
background in the publically available DES images. Using DES results
for non-central galaxies and our BCG profile, we obtain the total
stellar mass $M_{\star}=(20.1\pm3.7)\times10^{11}\,\Msun$,
$(31.6\pm5.8)\times10^{11}\,\Msun$, and
$(40.8\pm6.4)\times10^{11}\,\Msun$ at $r=0.5$, $1$, and $1.5\,\r500$,
respectively, in good agreement with our values reported in
Table~\ref{tab:masses}.

\bibliography{references}{}
\bibliographystyle{aasjournal}

\end{document}